\documentclass[usenatbib]{mn2e}

\usepackage[dvips]{graphicx}
\usepackage{amssymb}
\usepackage{txfonts}

\newcommand{\integral}{{\textit{INTEGRAL}}}
\newcommand{\xte}{{\textit{RXTE}}}

\newcommand{\sax}{{\textit{Beppo\-SAX}}}
\newcommand{\gro}{{\textit{CGRO}}}
\newcommand{\fermi}{{\textit{Fermi}}}
\newcommand{\agile}{{\textit{AGILE}}}
\newcommand{\swift}{{\textit{Swift}}}
\newcommand{\suzaku}{{\textit{Suzaku}}}
\newcommand{\msun}{{\rm M}_{\sun}}

\newcommand{\g}{$\gamma$}

\let\oldhat\hat

\renewcommand{\hat}[1]{\oldhat{\mathbf{#1}}}

\newbox\grsign \setbox\grsign=\hbox{$>$} \newdimen\grdimen \grdimen=\ht\grsign
\newbox\simlessbox \newbox\simgreatbox \newbox\simpropbox
\setbox\simgreatbox=\hbox{\raise.5ex\hbox{$>$}\llap
     {\lower.5ex\hbox{$\sim$}}}\ht1=\grdimen\dp1=0pt
\setbox\simlessbox=\hbox{\raise.5ex\hbox{$<$}\llap
     {\lower.5ex\hbox{$\sim$}}}\ht2=\grdimen\dp2=0pt
\setbox\simpropbox=\hbox{\raise.5ex\hbox{$\propto$}\llap
     {\lower.5ex\hbox{$\sim$}}}\ht2=\grdimen\dp2=0pt

\title[Orbital modulation of X-rays in Cyg X-3]{Energy-dependent orbital modulation of X-rays and constraints on emission of the jet in Cyg X-3}

\author[A. A. Zdziarski et al.]
{Andrzej A. Zdziarski,$^{1}$\thanks{E-mail: aaz@camk.edu.pl} Chandreyee Maitra,$^{2,3}$ Adam Frankowski$^{1,4}$,
\newauthor Gerald K. Skinner$^{5,6}$ and Ranjeev Misra$^7$\\
$^1$Centrum Astronomiczne im.\ M. Kopernika, Bartycka 18, PL-00-716 Warszawa, Poland\\
$^2$Raman Research Institute, Sadashivnagar, Bangalore 560080, India\\
$^3$Joint Astronomy Programme, Indian Institute of Science, Bangalore 560012, India\\
$^4$ISDC Data Centre for Astrophysics, University of Geneva, ch.\ d'Ecogia 16,  CH-1290 Versoix,  Switzerland\\
$^5$Astroparticle Physics Laboratory, Code 661, CRESST and NASA/Goddard Space Flight Center, Greenbelt, MD 20771, USA\\ 
$^6$Department of Astronomy, University of Maryland, College Park, MD 20742, USA\\
$^7$Inter University Centre for Astronomy and Astrophysics, Pune University Campus, Pune 411007, India\\
}

\date{Accepted 2012 June 29.  Received 2012 June 28; in original form 2012 May 20}

\pagerange{\pageref{firstpage}--\pageref{lastpage}}
\pubyear{2012}

\begin{document}

\maketitle

\label{firstpage}

\begin{abstract}
We study orbital modulation of X-rays from Cyg X-3, using data from \swift, \integral\/ and \xte. Using the wealth of the presently available data and an improved averaging method, we obtain energy-dependent folded and averaged light curves with unprecedented accuracy. We find that above $\sim$5 keV, the modulation depth decreases with the increasing energy, which is consistent with the modulation being caused by both bound-free absorption and Compton scattering in the stellar wind of the donor, with minima corresponding to the highest optical depth, which occurs around the superior conjunction. We find a decrease of the depth below $\sim$3 keV, which appears to be due to re-emission of the absorbed continuum by the wind in soft X-ray lines. Based on the shape of the folded light curves, any X-ray contribution from the jet in Cyg X-3, which emits \g-rays detected at energies $>0.1$ GeV in soft spectral states, is found to be minor up to $\sim$100 keV. This implies the presence of a rather sharp low-energy break in the jet MeV-range spectrum. We also calculate phase-resolved \xte\/ X-ray spectra, and show the difference between the spectra corresponding to phases around the superior and inferior conjunctions can indeed be accounted for by a combined effect of bound-free absorption in an ionized medium and Compton scattering. 
\end{abstract}

\section{Introduction}
\label{intro}

Cyg X-3 is a high-mass X-ray binary with a Wolf-Rayet (WR) donor \citep{v92,v96,v93} and a very short orbital period of $P\simeq 4.8$ h. Its distance is $\simeq\! 7$--9 kpc \citep*{lzt09,d83,p00}. In spite of its discovery already in 1966 \citep{giacconi67}, Cyg X-3 remains poorly understood. In particular, the nature of its compact object remains uncertain, due to the lack of reliable determination of the mass functions and inclination (see, e.g., \citealt{v09} for a discussion). However, the presence of a black hole is favoured by considering the X-ray and radio emission and the bolometric luminosity (\citealt{h08,h09,sz08},hereafter SZ08, \citealt*{szm08}, hereafter SZM08). Also, \citet*{zmg10} have shown that the differences in the form of the X-ray spectra of Cyg X-3 from those of confirmed black-hole binaries can be accounted for by Compton scattering in a cloud formed by the stellar wind from the companion. 

Cyg X-3 is a persistent X-ray source with a typical X-ray luminosity of $L_{\rm X}\sim 10^{38}$ erg s$^{-1}$. Its X-ray spectra have been classified into five states by SZM08, who have also quantified their correlations with the radio emission. Its high-energy \g-ray emission has been discovered by the \fermi\/ Large Area Telescope (LAT) and by \agile\/ in the soft spectral states (\citealt{fermi09}, hereafter FLC09; \citealt{agile}). Later detections by the LAT and \agile\/ are presented by \citet{williams11}, \citet{corbel12} and \citet{bulgarelli12}, hereafter B12.

Cyg X-3 shows pronounced flux modulation on the 4.8 h period, discovered in X-rays \citep{parsignault72,sh72,canizares73} and in infrared \citep{becklin73}. Both periodicities are most likely related to the strong stellar wind from the WR star (e.g., \citealt*{wkp85,v93}). In particular, the X-ray flux minima correspond to the maximum absorption in the line of sight around the superior conjunction (the compact object behind the WR star). The X-ray period has been found to increase \citep*{mmr78,ldf79,elsner80,vb81,vb89,kitamoto87,kitamoto95,singh02}. The period increase of $\dot P/P\simeq 1\times 10^{-6}$ yr$^{-1}$ is likely to be due to the loss of angular momentum through the wind mass loss \citep{do74}.

In this work, we present a detailed study of the dependence of the form of the X-ray orbital modulation on energy, using data from \swift, \integral\/ and {\it Rossi X-ray Timing Explorer\/} (\xte). We also study the dependence of the modulation on the X-ray spectral state, as well as obtaining X-ray modulation profiles for intervals simultaneous with the detections of high-energy \g-rays. The \g-ray emission was found to be also strongly modulated on the orbital period, but with the maxima and minima approximately corresponding to the X-ray minima and maxima, respectively (FLC09). This phase reversal has been interpreted as due to anisotropy of Compton scattering of the stellar emission by relativistic electrons in the jet \citep*{dch10}. The jet emission certainly extends to lower energies, and, using the model of \citet{dch10}, the hard X-rays from the jet are also predicted to have the maximum around the superior conjunction (\citealt{z12}, hereafter Z12), i.e., shifted with respect to the observed X-ray modulation by a half of the period. A measurement of the energy-dependent orbital modulation up to hard X-rays simultaneous with the \g-ray emission can then constrain the X-ray contribution of the jet.

\section{Folded light curves from X-ray monitoring}
\label{monitoring}

\subsection{The X-ray monitoring data}
\label{data}

\begin{figure}
\centerline{\includegraphics[width=\columnwidth]{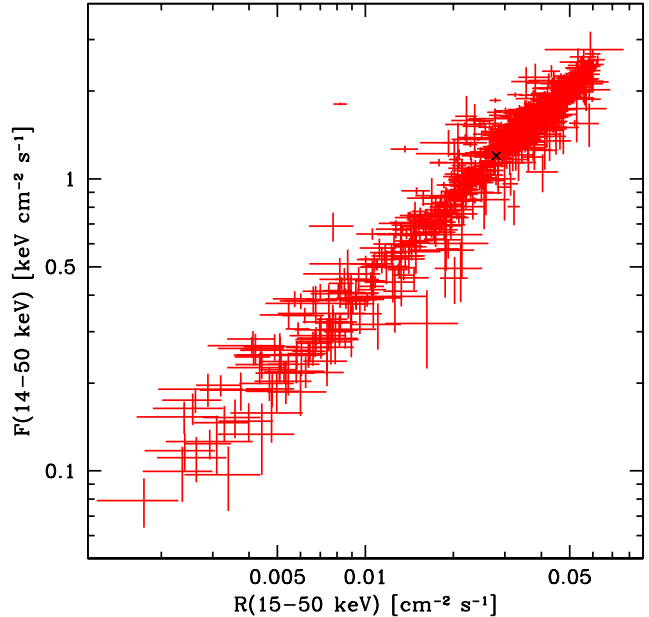}} 
\caption{The correlation between the daily averages of the BAT flux, $F(14$--50 keV) from our analysis vs.\ those of the 15--50 keV rate in the public BAT data, see Section \ref{data}. Only points with statistical significance $>3\sigma$ (for each of the quantities) are shown. The black cross shows the flux/rate  
of $F(14$--50 keV$)=1.2$ keV cm$^{-2}$ s$^{-1}$, corresponding to $R(15$--50 keV$)=0.028$ cm$^{-2}$ s$^{-1}$, below which we define our soft spectral state. 
} \label{bat1_bat8}
\end{figure}

\begin{figure}
\centerline{\includegraphics[width=\columnwidth]{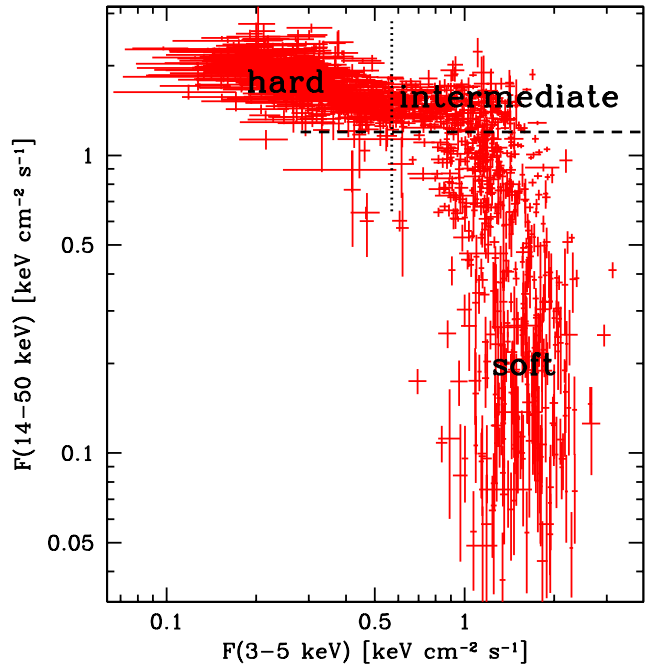}} 
\caption{The correlation between the daily-averaged energy fluxes in the 3--5 keV and 14--50 keV ranges. The hard and soft states are defined here by $F(3$--5 keV$)< 0.58$ keV cm$^{-2}$ s$^{-1}$ and $F(14$--50 keV$)< 1.2$ keV cm$^{-2}$ s$^{-1}$, respectively. The boundaries are marked by the dotted and dashed lines. An intermediate state corresponds to both the 3--5 keV and 14--50 keV fluxes above the respective boundaries. Only points with statistical significance $>3\sigma$ (for each of the fluxes) are shown.
} \label{correlation}
\end{figure}

We use X-ray monitoring data from the \swift\/ Burst Alert Telescope (BAT; \citealt{barthelmy05,m05}) in the form of a 14--195 keV 8-channel light curve created for this study. The typical exposure for a single data point is $\sim 10^3$ s. The data contain 23783 points within 1807 days spanning MJD 53355--55469. The channels are between energies of 14, 20, 24, 35, 50, 75, 100, 150 and 195 keV. These data have been obtained using an analysis different from that of http://swift.gsfc.nasa.gov/docs/swift/results/transients, which gives BAT count rates in the single 15--50 keV channel. Fig.\ \ref{bat1_bat8} shows a comparison of the daily-averaged 15--50 keV count rates with the corresponding 14--50 keV energy fluxes from our analysis. The energy fluxes have been obtained by converting the count rates of our data using scaling to the Crab spectrum (see \citealt*{zps11}). 

We also use both dwell and daily-averaged data from the All-Sky Monitor (ASM; \citealt*{brs93,levine96}) on board \xte. The dwell data contain 97996 measurements within 5267 days spanning MJD 50087--55915, and the exposure of a single observation is $\la 90$ s. The ASM has three channels at energies of 1.5--3 keV, 3--5 keV and 5--12 keV, for which the corresponding energy fluxes are obtained by scaling to the Crab spectrum, as above. 

Fig.\ \ref{correlation} shows the relationship between the energy fluxes in the 3--5 keV and 14--50 keV photon energy ranges. We see that at low soft X-ray fluxes, about $F(3$--5 keV$)< 0.58$ keV cm$^{-2}$ s$^{-1}$ (marked by the dotted line), corresponding to the ASM 3--5 keV count rate $< 3$ s$^{-1}$, there is a clear soft/hard X-ray flux anticorrelation. Note that a positive correlation between the 3--5 keV count rate and radio flux is seen below the same rate (SZM08). At higher 3--5 keV fluxes but at high 14--50 keV fluxes ($F(14$--50 keV$)\ga 1.2$ keV cm$^{-2}$ s$^{-1}$, marked by the dashed line), the hard X-ray flux changes within a narrow range independent of the soft X-ray flux. At lower 14--50 keV fluxes (below the dashed line), there is an apparent anticorrelation down to $F(14$--50 keV$)\sim 0.5$ keV cm$^{-2}$ s$^{-1}$. Below it, the soft X-ray flux changes within a narrow range approximately independent of the hard X-ray flux. Such anticorrelations, expressing a spectral pivoting around $\sim 10$--15 keV, can also be inferred from figs.\ 1--2 of SZM08. Fig.\ 7b of SZM08 and fig.\ 3 of \citet{h08} show similar correlations with the 20--100 keV flux measured (though with a lower sensitivity than that of the BAT) by the Burst and Transient Source Experiment (BATSE) on board {\it Compton Gamma Ray Observatory\/} (\gro).

We have also studied MAXI \citep{matsuoka09} data. However, the exposure of a single data point for those data is 1.5 h, which is $\sim P/3$ and much longer than any orbital bin we use. Thus, we do not use those data for studying orbital modulation in Cyg X-3. We note those data also provide us with a hard-state condition, of the 2--4 keV photon flux $<0.2$ cm$^{-2}$ s$^{-1}$, approximately equivalent to the ASM 3--5 keV count rate $<3$ s$^{-1}$.

We performed analyses including all of the ASM data and similarly with all of the BAT data, as well as some subsets of each. First, we use a hard-state criterion of the daily-averaged 3--5 keV ASM count rate $<3.0$ s$^{-1}$, corresponding to $F(3$--5 keV$)< 0.58$ keV cm$^{-2}$ s$^{-1}$, to the left of the dotted line in Fig.\ \ref{correlation}. As the soft-state criterion, we use the daily-averaged 14--50 keV BAT flux of $< 1.2$ keV cm$^{-2}$ s$^{-1}$, i.e., below the dashed line in Fig.\ \ref{correlation}. Our intermediate state corresponds to both fluxes being above these boundaries, i.e., in the upper right corner of Fig.\ \ref{correlation}. The soft-state criterion corresponds to the intervals of MJD 53746--53896, 53913--54138, 54204--54252, 54445--54470, 54548--54642, 54748--54818, 54980--55044, 55080--55103, 55322--55349 and 55584--55645 (determined using a 5-d running average in order to reduce the effect of fluctuations). For epochs after the end day of our BAT data, we have applied the corresponding criterion for the public 15--50 keV BAT data.

We also use data from intervals corresponding to the detected high-energy \g-ray emission. Currently, they are MJD 54566--54647 (B12), 54750--54820 (FLC09), 54821--54850 (B12), 54990--55045 (FLC09), 55324--55326 (B12), 
55343--55345 (\citealt{williams11}; B12) and 55586--55610, 55642-55644 \citep{corbel12}. All those periods correspond to the soft state as defined above, i.e., below the dashed line in Fig.\ \ref{correlation}, see FLC09 and B12. Thus, the appearance of \g-ray emission corresponds to a marked change of the form of the soft/hard X-ray correlation. Combining the results from \fermi\/ LAT and \agile, it appears likely that all low hard-flux intervals are also associated with some high-energy \g-ray emission.

\subsection{Treatment of the light curves}
\label{models}

Cyg X-3 has an orbital period that is increasing with time. Thus, we use a quadratic ephemeris,
\begin{equation}
T_m=T_0+P_0 m +c m^2,\quad c=P_0 \dot P/2,\quad P=P_0 + 2c m,
\label{ephemeris}
\end{equation}
where $T_m$ is the time of a $m$-th occurrence of a zero orbital phase (presumably related to the superior conjunction) since the reference time, $T_0$, $P_0$ is the period at $T_0$, $\dot P$ is the period derivative, and $P$ is the period at $T_m$. The ephemeris (in UTC) of \citet{singh02} is
\begin{equation}
T_0=40949.392\, [{\rm MJD}], \, P_0=0.19968443\,{\rm d},\, c=5.75\times 10^{-11}\,{\rm d}.
\label{singh}
\end{equation}
An updated ephemeris taking into account \suzaku\/ observations (S. Kitamoto, private communication) is
\begin{equation}
T_0=40949.3913\, [{\rm MJD}], \, P_0=0.19968451\,{\rm d},\, c=5.62\times 10^{-11}\,{\rm d},
\label{kitamoto}
\end{equation}
in which case $\dot P/P\simeq (1.03\pm 0.02)\times 10^{-6}$ yr$^{-1}$.

We note that these ephemerides of Cyg X-3 use the template of \citet{vb89}, which is defined numerically by their table 2. That template has the minimum at the phase $\simeq 0.96$--0.97 instead of 1.0 (or, equivalently, phase 0.0). Thus, minima at phases $<1$ obtained by us do not indicate a discrepancy with respect to previous results. We also correct the light curves for barycentric delays. 

We need to properly average the folded light curves, [$F_i(t_i)$, $\sigma_i(t_i)$], where $F_i$ is the average count rate of the observation with the mid-time of $t_i$, and $\sigma_i$ is its measurement error. One issue to consider is the large dynamic range of the flux variability of Cyg X-3, which is almost two orders of magnitude (see Fig.\ \ref{correlation}). Thus, linearly-averaged modulation profiles are strongly biased by those at highest flux states. Using flux logarithms can, in principle, somewhat alleviate this problem, as it strongly reduces the dynamic range used in summation. However, this requires removal of all negative flux measurements. For a weak signal, this introduces a strong bias against the minima of the modulation. We found that this effect strongly suppresses the depth of the modulation at energies $\ga 30$ keV. 

On the other hand, orbital modulation acts on a signal with a local flux level, the level being governed by the source aperiodic variability. The periodic and aperiodic variabilities are largely independent of each other. Thus, in order to effectively reduce the large amplitude of the underlying aperiodic flux variability, we calculate a running average, averaging all observations within some $\pm \Delta t$ of a given observation time,
\begin{equation}
\bar F_i={1\over J}\sum_{j} F_j,\quad \vert t_i-t_j\vert \leq \Delta t,
\label{running}
\end{equation}
where $J$ is the number of data points satisfying the above condition on $j$. Then, we renormalize both $F_i$ and $\sigma_i$ by dividing each by $\bar F_i$ (so $\sigma_i/F_i$ remain unchanged). In this way, effects of the aperiodic long-term variability on time scales $\ga 2\Delta t$ are removed. In the case of measurements equally spaced at a time interval of $\ll P/K$, where $K$ is the number of bins per period at which the folded light curve is calculated, using $\Delta t=P/2$ would optimally remove the effect of long-term aperiodic variability. However, our measurements do not satisfy this criterion, and are occasionally sparsely and unevenly spaced. Then, the shorter the $\Delta t$, the worse the statistics on which the running average is based. In the extreme case of no other measurements within $\pm \Delta t$ from a given observation time, $F_i$ would be reset to unity, and such a point would falsely reduce the actual orbital modulation. Thus, in our method, we reject points for which the running average is based on $J<10$ points. We have tested this method for our data, and have found that $\Delta t= 1$ d ($\simeq 5 P$) is approximately the shortest interval for which the running average is estimated with good statistics and only a small fraction of points have $J<10$. Namely, $\langle J\rangle\pm {\rm rms}$, and the fraction of the rejected points are ($51\pm 23$, 0.02), ($34\pm 16$, 0.04), for the ASM and BAT data, respectively. We thus use the above value of $\Delta t$ hereafter.

Another issue here is related to aperiodic short time-scale variability on time scales shorter than the length of an orbital bin, which effect we also would like to remove. Therefore, we pre-average the light curves on real-time intervals with the length equal to the bin size, $P/K$, where $K$ is the number of phase bins per orbit. We obtain local averages of the renormalized (see above) flux and its error, which we denote as $F_{kl}$, $\sigma_{kl}$, respectively, where $k$ is the number of the phase bin and $l$ is the number of the time bin (counted from the start time of a given data set) contributing to the $k$-th phase bin. This method was applied in studying periodic variability of Cyg X-1 by \citet*{izp07}. It also partly removes a bias on folded averages due to a non-uniform coverage, e.g., due to a large number of points during one time interval and a low number of points during another, with both contributing to the same phase bin, $k$. Without pre-averaging, the former would have much higher weight than the latter, even if all measurement were relatively accurate. The pre-averaging does not have a major effect in the case of Cyg X-3 due to its very short orbital period (resulting in rather few points per a $P/K$ time interval), as well as for ASM and BAT data, which have relatively uniform coverage. Still, this is the statistically correct procedure, resulting, in particular, in correctly estimated uncertainties of the folded and averaged light curves.

In final averaging, we use weights given by inverse squares of the uncertainties, 
\begin{equation}
\langle F_k\rangle = {\sum_{l=1}^{N_k} F_{kl}/\sigma_{kl}^2\over \sum_{l=1}^{N_k} 1/\sigma_{kl}^2},
\label{average}
\end{equation}
where $N_k$ is the number of time bins contributing to the $k$-th phase bin. The error of the weighted average is estimated from the variance, $(\langle F_k^2\rangle-\langle F_k\rangle^2)/(N_k-1)$,
\begin{equation}
\langle\sigma_k^2\rangle = {1\over N_k -1} \left({\sum_{l=1}^{N_k} F_{kl}^2/\sigma_{kl}^2\over \sum_{l=1}^{N_k} 1/\sigma_{kl}^2}-\langle F_k\rangle^2\right).
\label{sigma}
\end{equation}

We fit the obtained folded/averaged light curves by a simple model of absorption/scattering in a spherically symmetric stellar wind. A commonly-used wind velocity profile from a massive star (e.g., \citealt*{lcp87}) and the resulting electron density are, respectively,
\begin{equation}
v(r) \simeq v_{\infty}\left(1-{R_* \over r}\right)^{\beta},\quad n_{\rm e}(r) = {-\dot M \over 4 \upi m_{\rm p}\mu_{\rm e} r^2 v(r) },
\label{v_n}
\end{equation}
where $r$ is the distance from the centre of the donor of the radius $R_*$, $v_\infty$ is the terminal velocity, $\beta$ parametrizes the wind acceleration, $\dot M$ is the total mass-loss rate, $m_{\rm p}$ is the proton mass, $\mu_{\rm e}\simeq 2/(1+X)$ is the mean electron molecular weight, and $X$ is the H fraction. A small correction due to velocity reaching the sound speed rather than being null at the stellar surface has been neglected. 

Assuming the opacity for a given photon energy is a spatially independent constant times the density (as is the case, e.g., for Compton scattering or for bound-free absorption with a constant ionization coefficient), the optical depth is an integral over the photon path, $l$ \citep{pringle74},
\begin{equation}
\tau(\phi)= \tau_0 \int_0^\infty \left(r\over a\right)^{-2} \left(1-{R_*/a \over r/a}\right)^{-\beta} {\rm d}(l/a),
\label{tau_sc_v}
\end{equation}
where $\tau_0=(\sigma_{\rm C}+\sigma_{\rm ion}) n_0 a$, $\sigma_{\rm C}$ is the Compton cross section (including its Klein-Nishina decline), $\sigma_{\rm ion}$ is a photoionization cross section per electron, $n_0$ is a fiducial electron density at $r=a$ under the assumption of $v=v_\infty$ and $a$ is the orbital separation. For a circular orbit and assuming the radiation originating at the compact object location, $r$ is related to $l$ by
\begin{equation}
(r/a)^2=1+(l/a)^2- 2 (l/a) \sin i\cos\phi, 
\label{radius}
\end{equation}
where $i$ is the binary inclination and $\phi$ is the orbital phase (defined as $\phi=0$, $\upi$ at the superior and inferior conjunction, respectively), see fig.\ 1 in \citet{zdz12}. For $r\gg R_*$, i.e., $v\simeq v_\infty$, 
\begin{equation}
\tau(\phi)\simeq \tau_0 {\upi/2+\arcsin s\over \sqrt{1-s^2}},\quad s=\sin i\cos\phi,
\label{tau_sc}
\end{equation}
which is is equivalent to equation (8) of \citet{pringle74}.

In Compton scattering, a photon removed from one line of sight appears in a different one. Photons scattered away from directions with high $\tau$ preferentially leave the system close to the directions with low $\tau$. Also, the photon distribution will be a function of the angle with respect to the normal to the binary plane. It is possible to model this effect quantitatively in a more detailed treatment, e.g., using a Monte Carlo method. Such method was used by \citet*{hjr78}, who, however, presented their results for $i=90\degr$ only (and for isotropic wind). However, the wind density distribution in Cyg X-3 is certainly not isotropic, but instead with a strong focusing towards the compact object \citep{fc82}, and the wind structure is affected by irradiation from the X-ray source. Taking into account these complications is beyond the scope of this work. We therefore make a simplifying assumption (as in \citealt{pringle74}) that a scattering removes the photon from the line of sight and neglect photons scattered into it. This allows us to fit the observed phase-dependent fluxes by
\begin{equation}
F(\phi)/F_0 =\exp\left[-\tau(\phi-\phi_0)\right],
\label{flux}
\end{equation}
with $\tau_0$, $i$, $\beta$, the offset phase, $\phi_0$, and $F_0$ being the model free parameters. We note that $F_0=F_{\rm max}\exp[\tau(\upi)]$. This  requires the scattering optical depth to be $\la 1$ (though the absorption optical depth can be any). We define the fractional modulation depth as
\begin{equation}
D\equiv {F_{\rm max}-F_{\rm min}\over F_{\rm max}}=1-\exp(-\Delta \tau), \quad \Delta\tau\equiv \tau(0)-\tau(\upi),
\label{eq:depth}
\end{equation}
where $F_{\rm max}$ and $F_{\rm min}$ are the maximum and minimum modulation fluxes, respectively. For the model of equation (\ref{tau_sc}), $\Delta\tau\simeq 2\tau_0 i/\cos i$.

We then need to specify the parameters of Cyg X-3 appearing in this model. Since Cyg X-3 is a He star, $X\simeq 0$ and $\mu_{\rm e}\simeq 2$. The stellar radius, $R_*$, is closely related to the WR mass, see fig.\ 2 of \citet{sm92}. The separation, $a$ follows from the total mass of the system. This, however, remains rather uncertain. Here, we choose, somewhat arbitrarily, a binary solution satisfying all the constraints of \citet{v09}, namely $M_*=20\msun$ and the compact object mass of $M_{\rm C}=6\msun$, for which $a\simeq 3.0\times 10^{11}$ cm. This $M_*$ corresponds to $R_*\simeq 1.0\times 10^{11}$ cm \citep{sm92}. This solution corresponds to an allowed range of inclinations of about $i\simeq (33\degr$--$63\degr)$, see section 6 of \citet{v09}. A typical value of the terminal wind velocity estimated for Cyg X-3 is $v_\infty\simeq 1.7\times 10^8$ cm s$^{-1}$ (see a discussion and references in SZ08). We assume $\beta=2$, which was found to fit well the results of wind particle simulations for Cyg X-3 of \citet{v09}. Also, this value was used by \citet{langer89} in his modelling of mass loss from WR stars. 

As an alternative, and purely phenomenological, model for the folded light curves, we use a sum of three Fourier harmonics (in the logarithmic space). We follow here exactly the treatment described in section 3.2 of \citet{lachowicz06}. The modulation depth in this model is determined numerically.

\subsection{Energy-dependent folded light curves}
\label{lc}

\begin{figure}
\centerline{\includegraphics[width=\columnwidth]{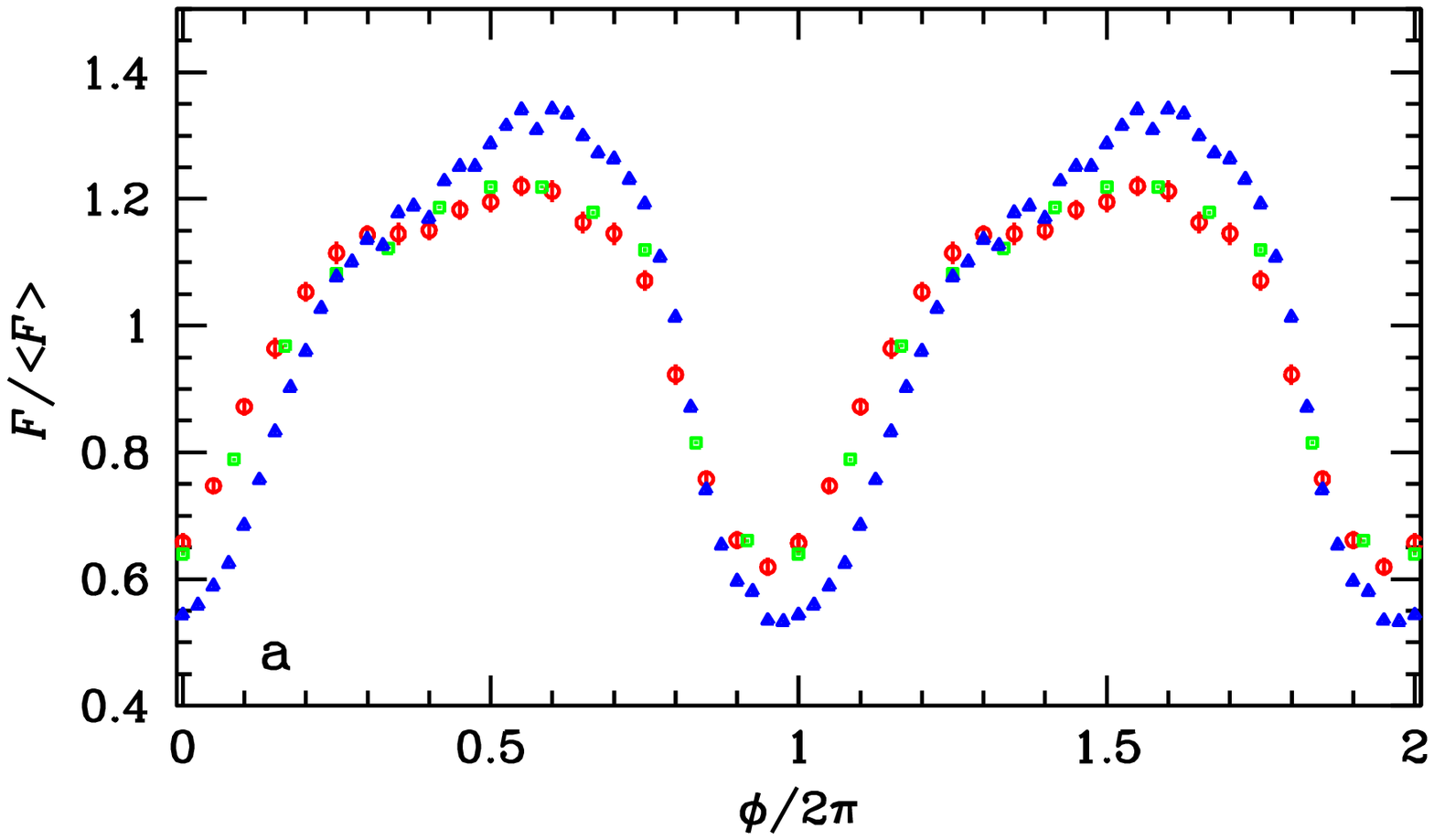}} 
\centerline{\includegraphics[width=\columnwidth]{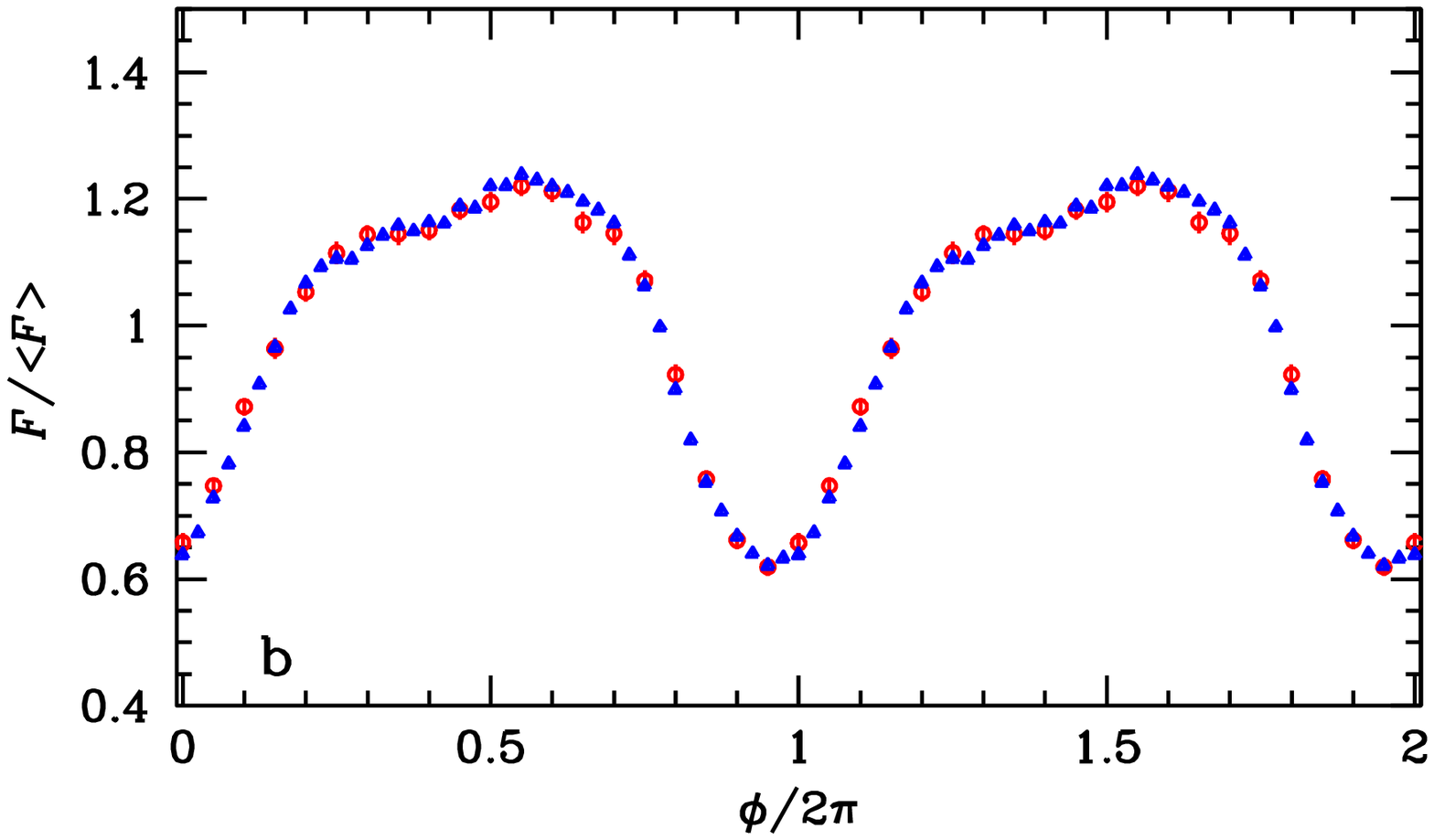}} 
\centerline{\includegraphics[width=\columnwidth]{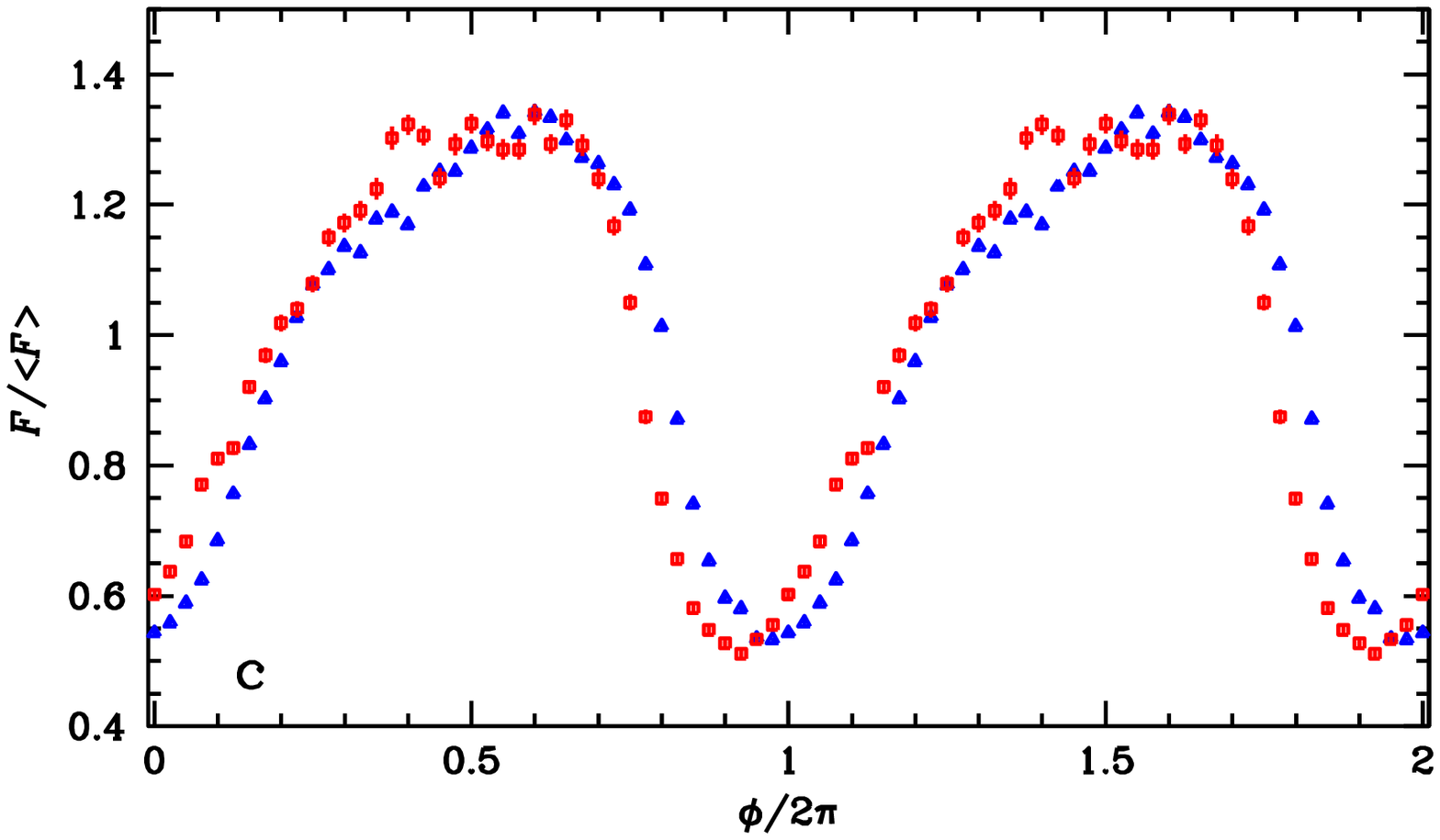}} 
\caption{(a) Folded/averaged light curves for the count rates for the energy range of 1.5--12 keV (blue, ASM), 14--50 keV (red, BAT) and 20--40 keV (green, \integral/ISGRI), using $K=40$, 20 and 12 bins, respectively. For the BAT data, only observations with the exposure $<0.5 P/K$ have been selected. Here and elsewhere, two orbital ranges are shown for clarity of the display. (b) Comparison of the BAT folded light curve with the above exposure criterion and $K=20$ with that without it (for $K=40$). (c) Comparison of the ASM folded light curve calculated with our treatment described in Section \ref{models} (blue triangles), i.e., using the barycentric-corrected data, the quadratic ephemeris of equation (\ref{kitamoto}), the renormalization of the light curve to the running average, and the pre-averaging, with that obtained using a constant orbital period, unweighted averaging and no other corrections (red squares).
} \label{asm_bat_integral}
\end{figure}

\begin{figure}
\centerline{\includegraphics[width=\columnwidth]{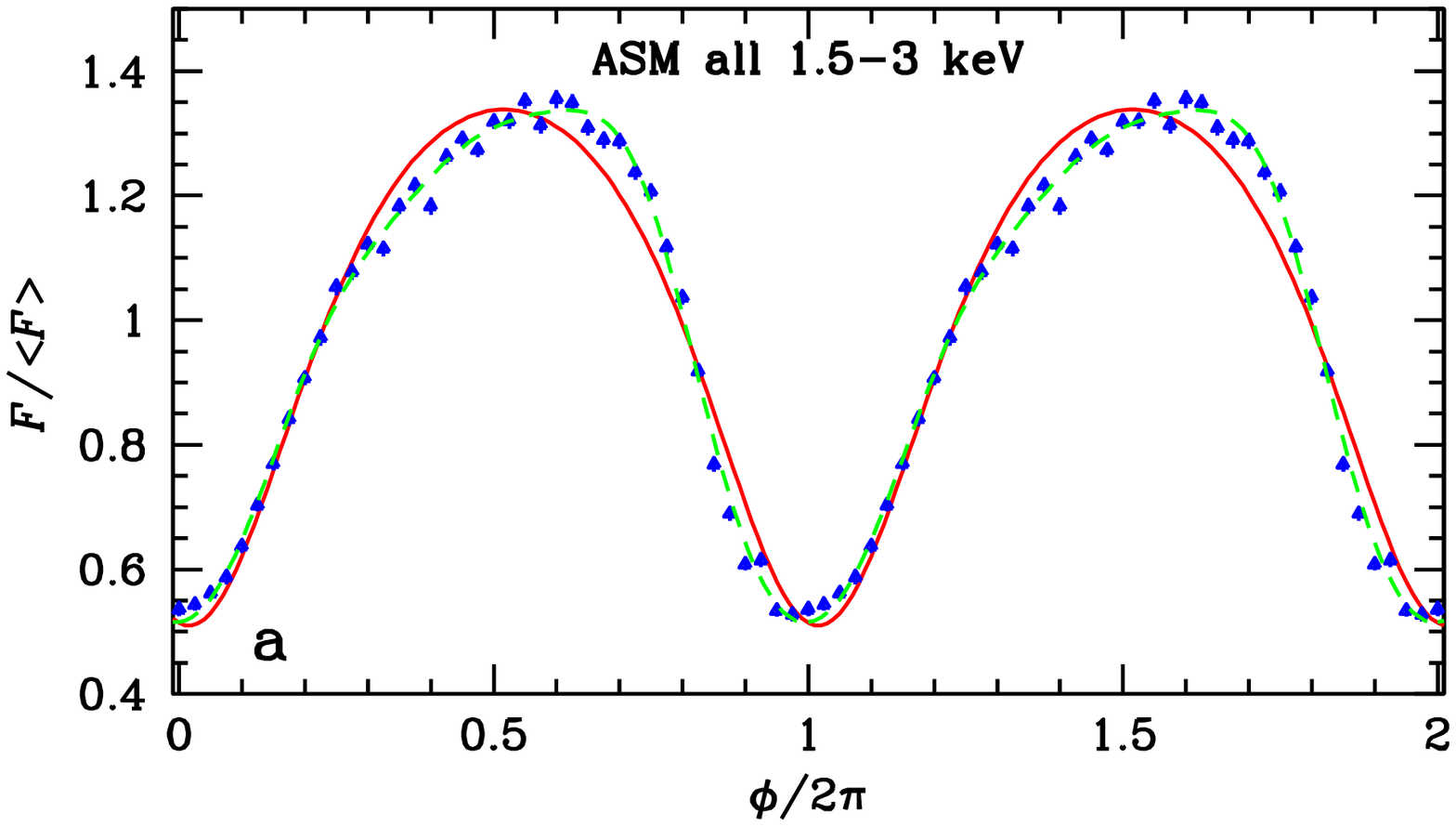}} 
\centerline{\includegraphics[width=\columnwidth]{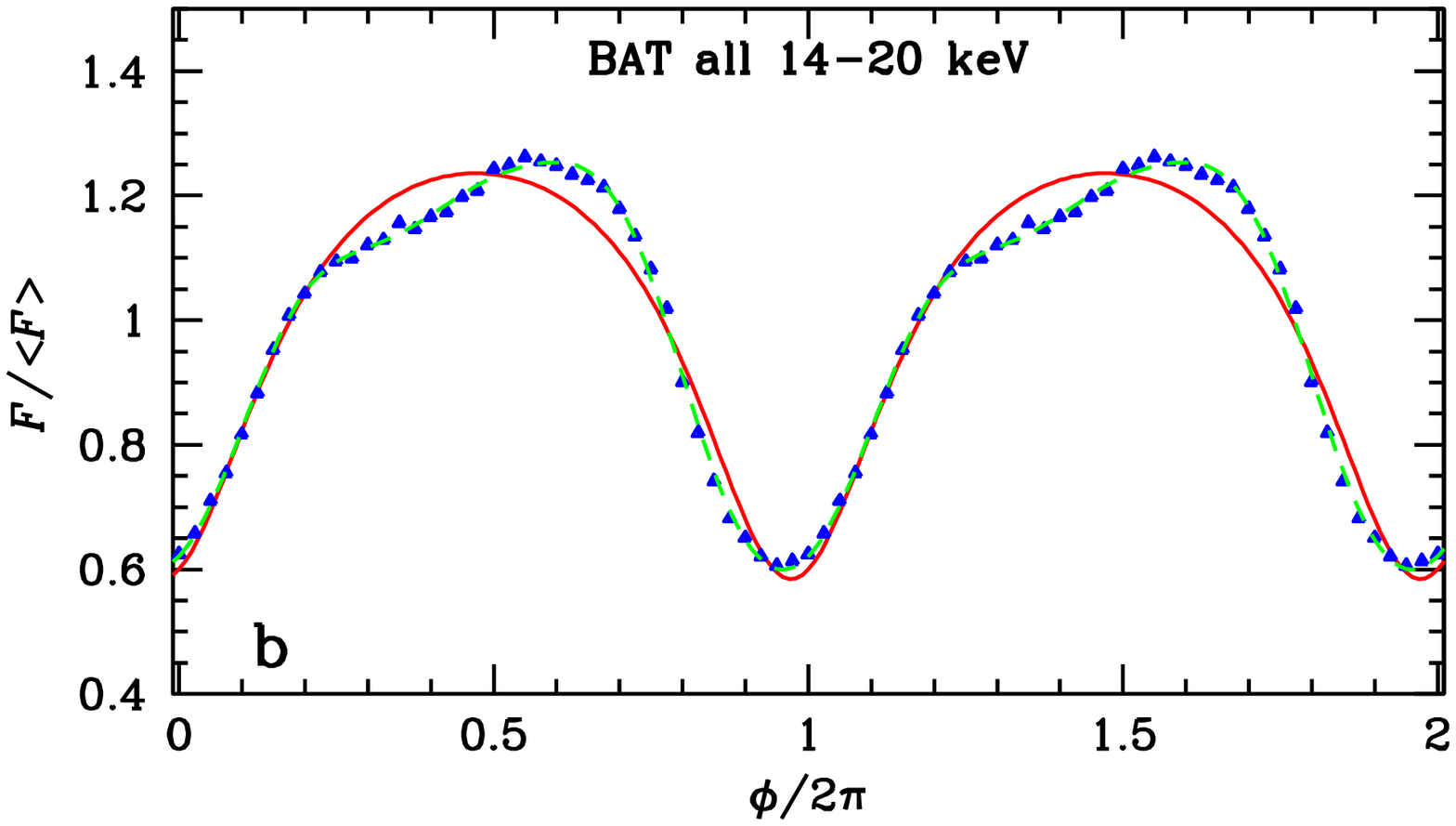}} 
\caption{Folded light curves using all of the data for (a) 1.5--3 keV ASM, and (b) 14--20 keV BAT data, using $K=40$. The red solid curves show fits to the data with the model of equation (\ref{tau_sc_v}), (\ref{flux}). The green dashed curves show fits by the 3-harmonic model.
} \label{asm_bat}
\end{figure}

\begin{figure}
\centerline{\includegraphics[width=\columnwidth]{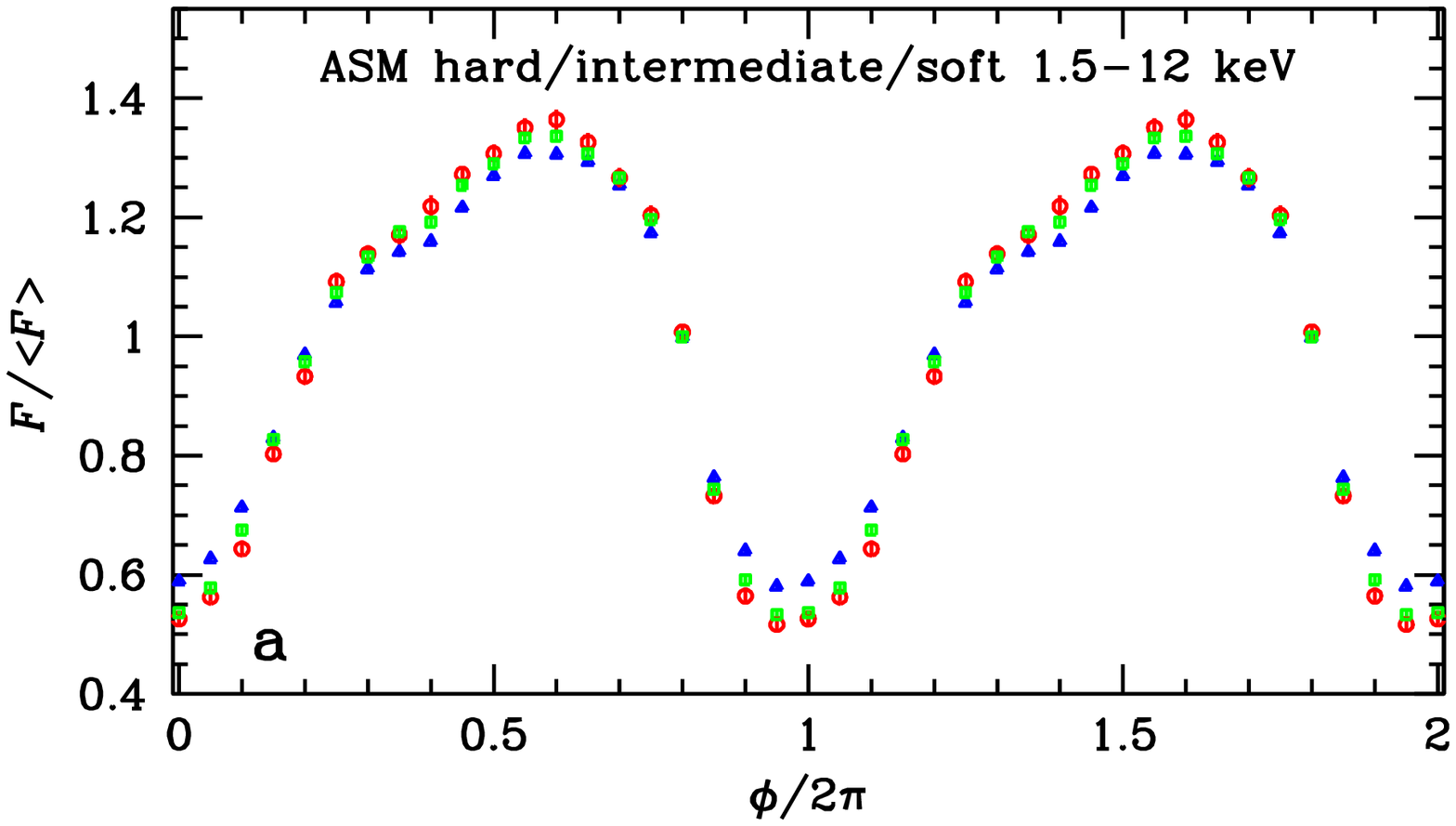}} 
\centerline{\includegraphics[width=\columnwidth]{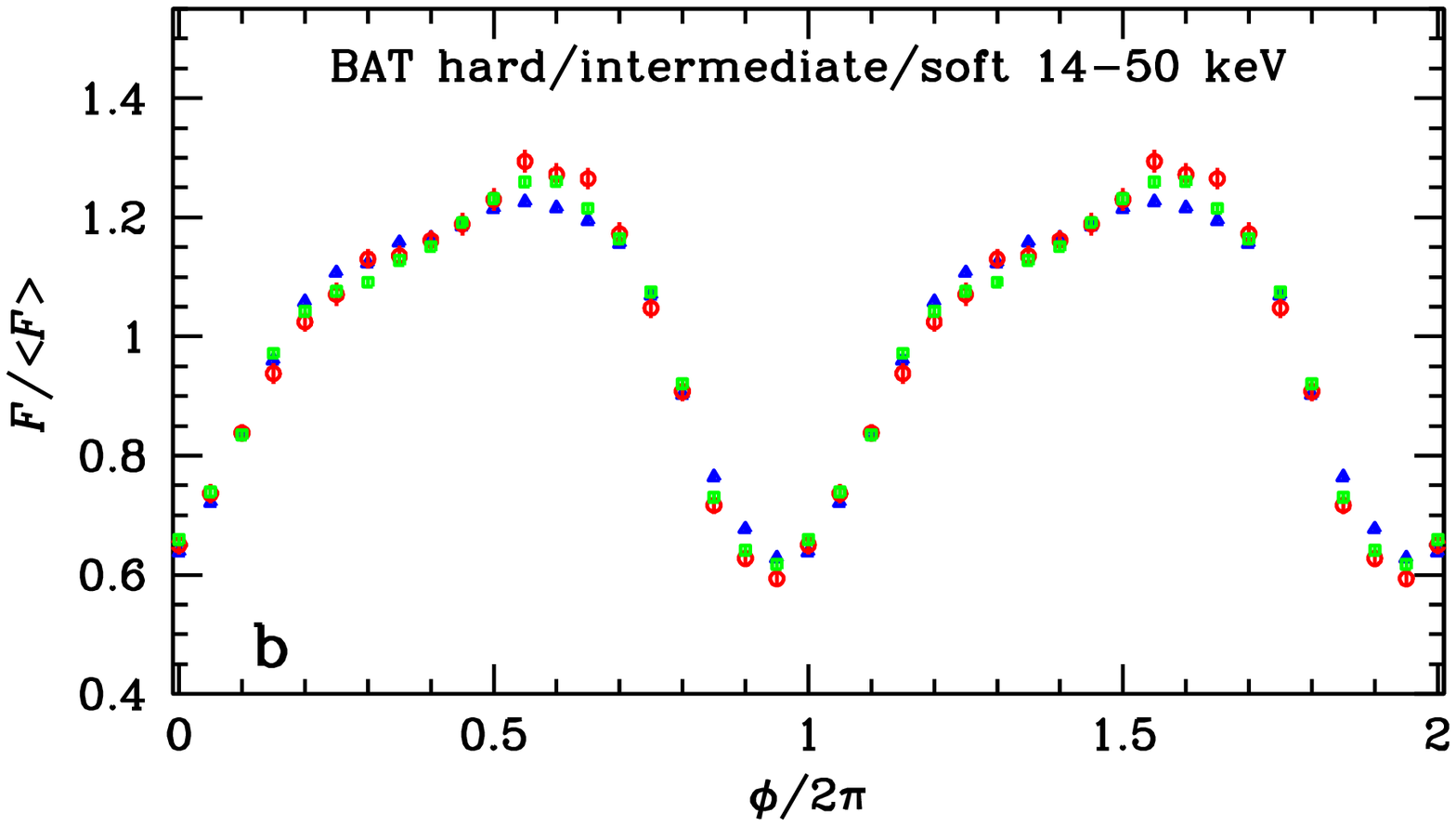}} 
\caption{Comparison of the folded light curves in the hard (blue triangles), intermediate (green squares) and soft (red circles) spectral states for (a) the ASM 1.5--12 keV range, and (b) the BAT 14--50 keV range. 
} \label{hard_soft}
\end{figure}

\begin{figure}
\centerline{\includegraphics[width=\columnwidth]{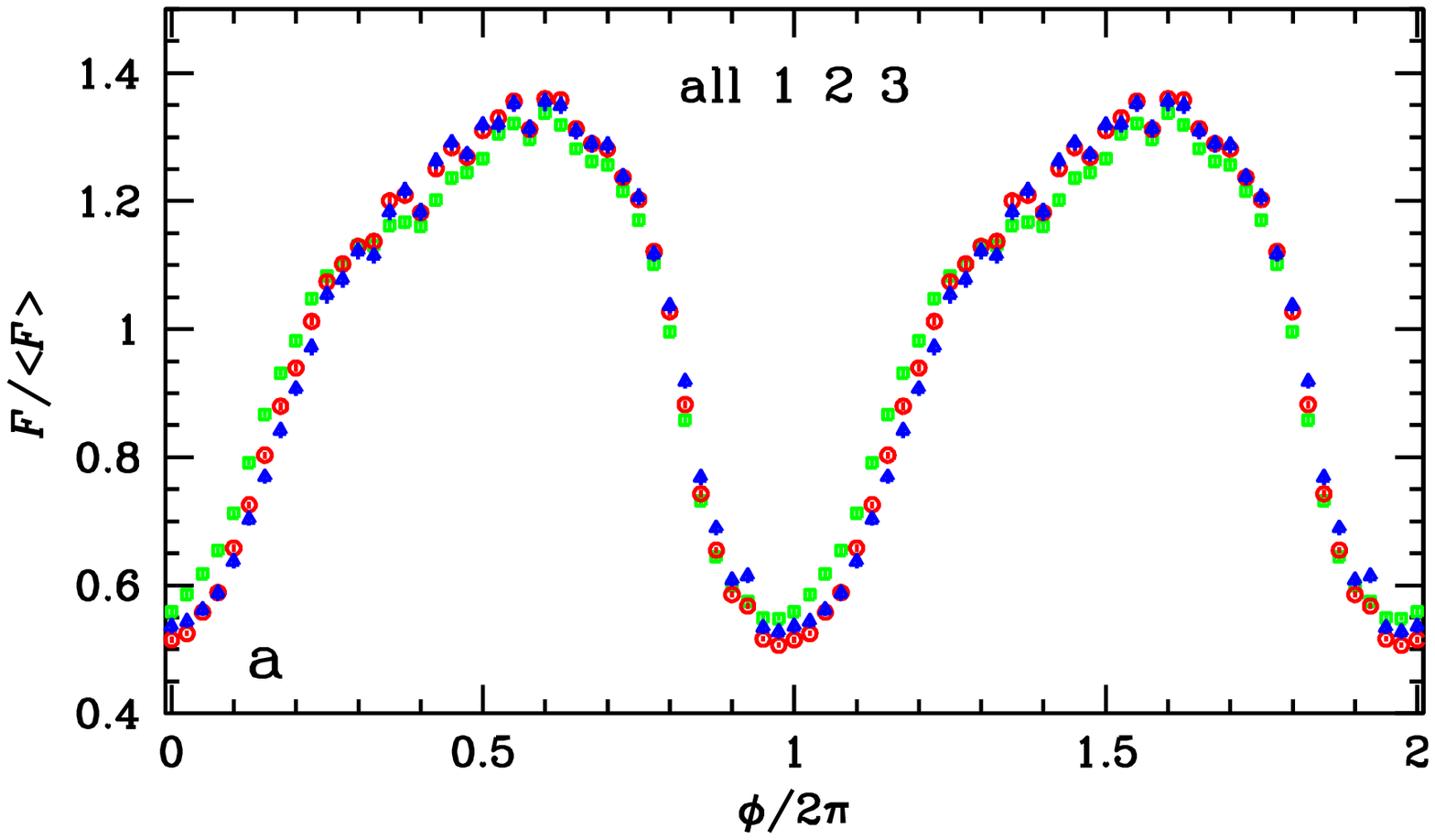}} 
\centerline{\includegraphics[width=\columnwidth]{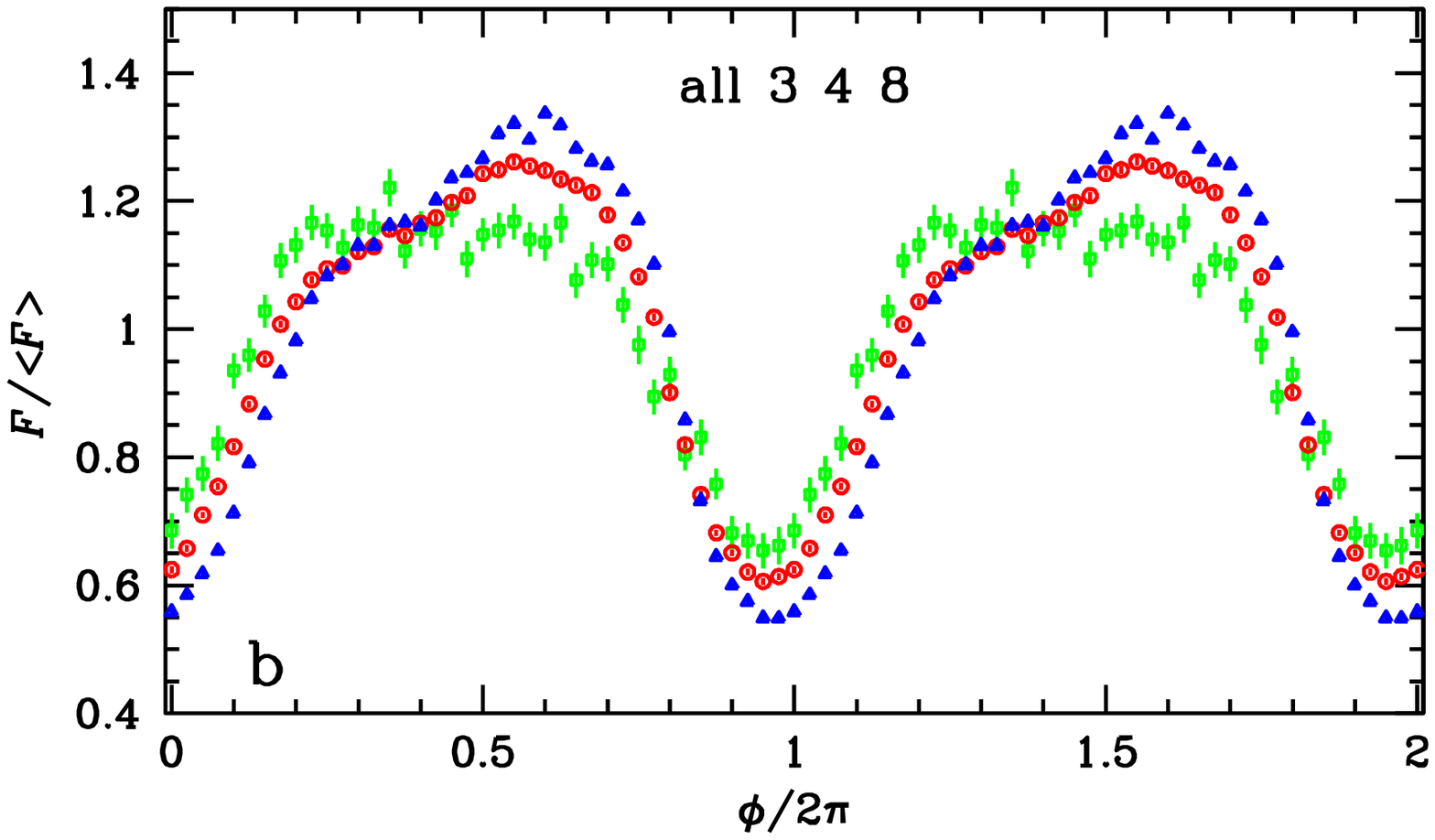}} 
\centerline{\includegraphics[width=\columnwidth]{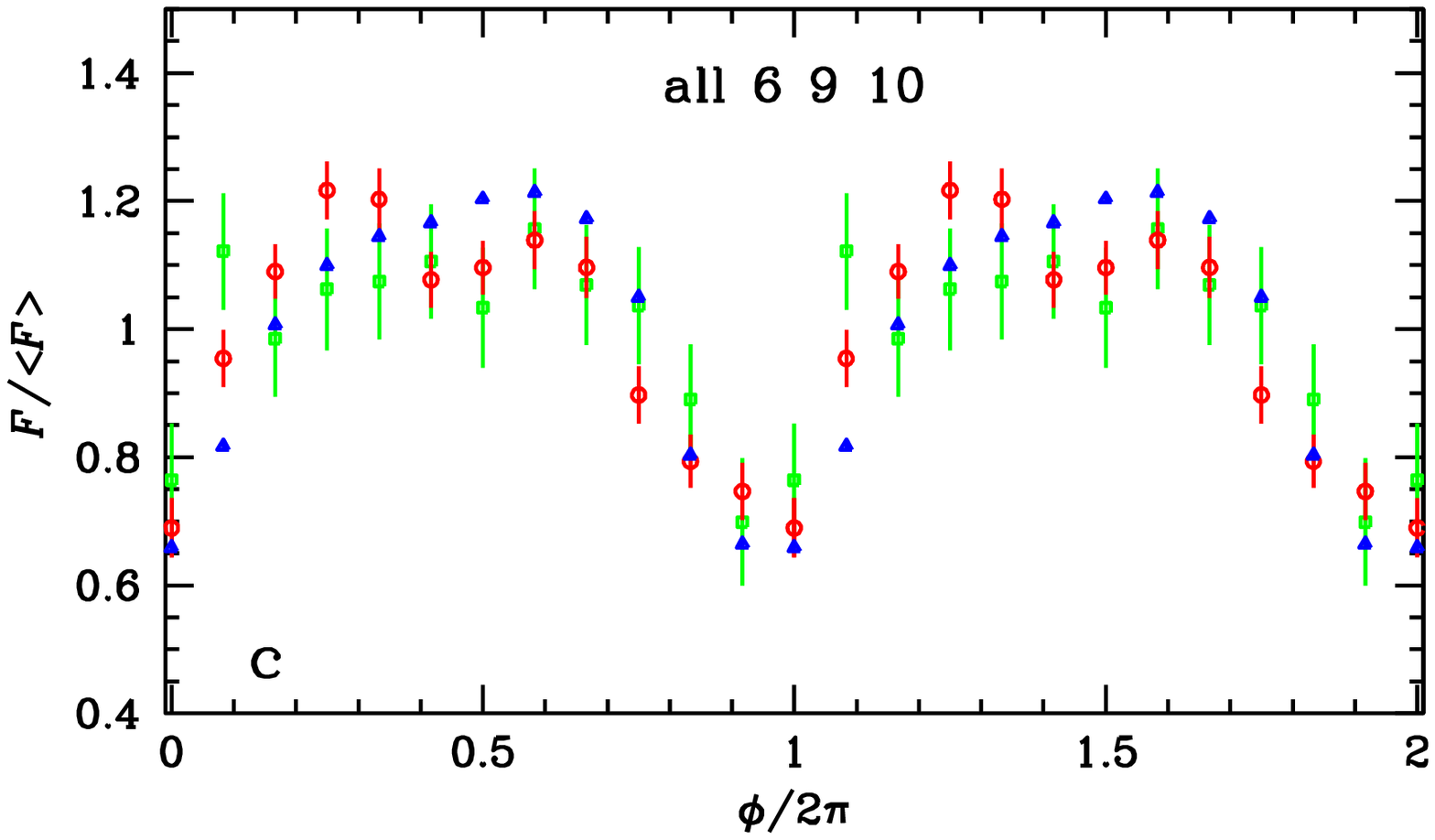}} 
\caption{Comparison of the folded light curves in different energy channels for all the available ASM and BAT data. The blue, red and green light curves are for channels (a) 1.5--3 keV, 3--5 keV, 5--12 keV, (b) 5--12 keV, 14--20 keV, 50--75 keV, and (c) 24--35 keV, 75--100 keV, 100--150 keV, respectively.
} \label{comparison}
\end{figure}

We first consider folded/averaged light curves averaged over all the available ASM and BAT data. We have compared the light curves folded with the ephemerides of equations (\ref{singh}) and (\ref{kitamoto}). We have found the differences are tiny, and we use the updated ephemeris of equation (\ref{kitamoto}) hereafter. 

Fig.\ \ref{asm_bat_integral}(a) shows the folded light curves for the total count rates of the ASM and BAT instruments (and also shows the 20--40 keV profile from \integral, see Section \ref{integral} below). We see the fractional modulation depth, equation (\ref{eq:depth}), is significantly larger in the 1.5--12 keV range, $D\simeq 0.61$, than the 14--50 keV one, $D \simeq 0.50$. This is likely due to the bound-free absorption by the wind being stronger in the ASM band than in the BAT one. The BAT minimum is at $\phi/2\upi\simeq 0.95$, somewhat less than the ASM one at $\simeq 0.97$. Both profiles have relatively complex and asymmetric shapes, with a flattening around the phase of 0.4 followed by a peak around 0.55-0.60. The ASM profile has two statistically significant dips, at the phases 0.40 and 0.575. For either of the dips, the difference with respect to the rate averaged over the two neighbouring bins is $>5$ standard deviations. Their presence indicates some complexity of the wind structure, probably due to focusing towards the compact object \citep{fc82}.  

An issue for the BAT data concerns their relatively long typical exposure of a single data point of $\sim 10^3$ s. This is comparable to the length of one orbital bin in the case of $K=20$, which may cause a smoothing of sharp features in the folded light curves. In Fig.\ \ref{asm_bat_integral}(a), we have thus used only measurements with the exposure $<0.5 P/K$ for the BAT profile. However, using only BAT data with short exposure times strongly reduces the number of used measurements. In the present case, accepting only exposures $<0.5 P/K$ for $K=20$ reduces the number of available measurements by a factor of $\sim 10$ with respect to using all the available data. In Fig.\ \ref{asm_bat_integral}(b), we compare the profiles with and without imposing this condition. We see that using the data without this selection causes certain smoothing of the profile, but the effect is rather minor. Thus, we do not use this selection for the BAT data hereafter. 

Fig.\ \ref{asm_bat_integral}(c) illustrates how taking into account various effects can affect folded/averaged light curves, for the example of the ASM light curve. The red squares show the profile obtained with a constant $P$ computed for the ASM data up to 2004 by \citet{wen06} (i.e., neglecting $\dot P$), neglecting the barycentric correction, without prior renormalizing with respect to the running average and without pre-averaging within time bins. We see that that curve has the minimum at $\phi/2\upi\simeq 0.9$ instead of 0.95 found for the treatment of Section \ref{models}. As we have checked, this shift is almost entirely due to neglecting the $\dot P$ of Cyg X-3. Taking into account barycentric time delays introduce only a very small effect. Furthermore, the red points have a much flatter and more scattered top part (similar to the profile shown in \citealt{wen06}). This difference is mostly due to renormalizing the light curve w/r to the running average. The profile without it is strongly dominated by the brightest states (as pointed out in Section \ref{models}). Then, fluctuations of those dominant states have a strong effect. On the other hand, our profile, shown by the blue triangles, is evenly averaged over all flux states. Finally, the pre-averaging introduces only a small effect for the ASM data, though it does noticeably increase the error bars.

The 14--50 keV BAT profile shown in Fig.\ \ref{asm_bat_integral} can be compared with the folded light curve in the 12--50 keV band from the Gamma-ray Burst Monitor (GBM) on board \fermi, shown in fig.\ 11 of \citet{wh12}. Unlike our, rather smooth, BAT profile, the GBM curve has multiple sharp peaks. We note that the orbital profile for Cyg X-1 shown in \citet{wh12} also has much stronger sharp features than the corresponding 15--50 keV BAT profile presented by \citet{zps11}. Possibly, the sharp GBM features are due to statistical fluctuations at somewhat underestimated measurement uncertainties. 

In Fig.\ \ref{asm_bat}(a--b), we fit the lowest-energy channels of the ASM and BAT with the model of equations (\ref{tau_sc_v}), (\ref{flux}). We see that neither profile is well fitted by this model. The profiles are significantly asymmetric, with a slower rise and a faster decline. (The scatter around the top of the 1.5--3 keV profile appears statistical.) Thus, the model does not provide statistically good fits, but we give the fit parameters for completeness: $i=33\degr\pm 1\degr$, $39\degr\pm 1\degr$, $\tau_0=0.27\pm 0.01$, $0.15\pm 0.01$, $\phi_0/2\upi=0.02\pm 0.01$, $-0.03\pm 0.01$, for the ASM and BAT, respectively (for the fixed $\beta=2$, see Section \ref{models}). Given the statistically unsatisfactory fits, these parameters may not correspond to the actual wind of Cyg X-3. Also, the pairs of the parameters ($\tau_0$, $\beta$) and ($i$, $\beta$) are each strongly anti-correlated. Still, our results indicate that, in spite of the strong X-ray orbital modulation, the inclination required to account for it does not need to be large, even for our spherically-symmetric wind model. The optical depths at the phases of 0, 0.25 and 0.5 are $\tau=(1.5$, 0.7, 0.5) and (1.0, 0.4, 0.3), respectively. Based on the fits, we can also calculate the corresponding mass loss rates using equation (\ref{v_n}), the definition of $\tau_0$, and the Cyg X-3 parameters adopted in Section \ref{models}. For the BAT and ASM best fits, $\tau_0$ corresponds to $-\dot M\simeq (0.8$--$1.4)\times 10^{-5}[\sigma_{\rm T}/(\sigma_{\rm C}+\sigma_{\rm ion})] \msun$ yr$^{-1}$ (where $\sigma_{\rm T}$ is the Thomson cross section), which is of the order of the values usually estimated for Cyg X-3 (see SZ08 and references therein). 

\begin{figure}
\centerline{\includegraphics[width=6.7cm]{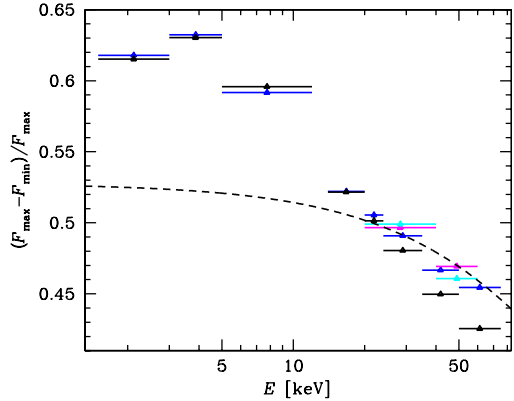}} 
\caption{The fractional modulation depth, $D(E)$, for all the available ASM and BAT data (blue and black symbols) and the ISGRI data (magenta and cyan symbols) estimated using fits with the model of equations (\ref{tau_sc_v}), (\ref{flux}) (blue and magenta) and with the 3-harmonic model (black and cyan). The dashed curve shows the dependence of $D(E)$ for equation (\ref{depth}) in the case of the Klein-Nishina cross section, approximately normalized to $D(20\,$keV).
} \label{depth}
\end{figure}

\begin{figure}
\centerline{\includegraphics[width=\columnwidth]{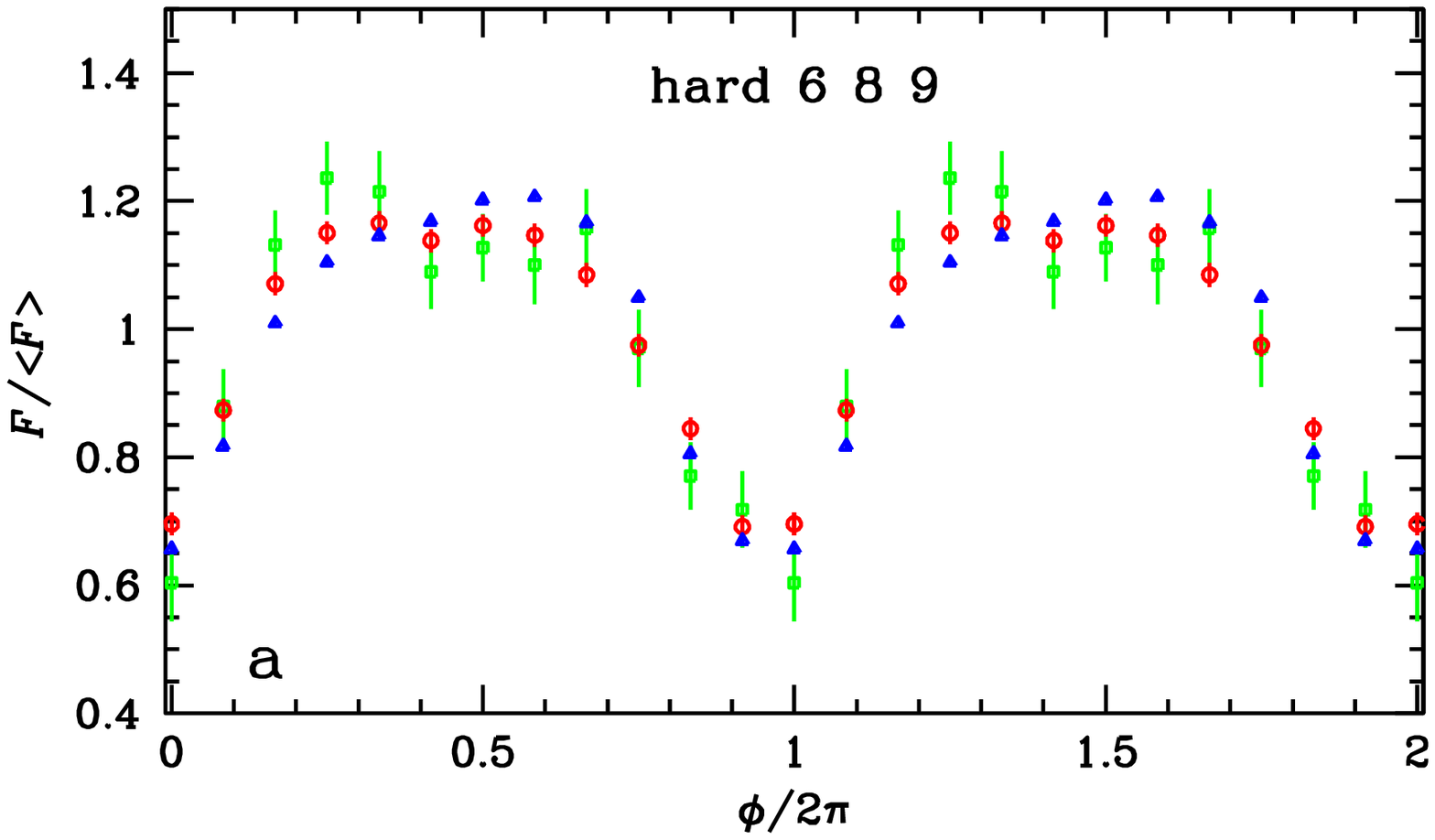}} 
\centerline{\includegraphics[width=\columnwidth]{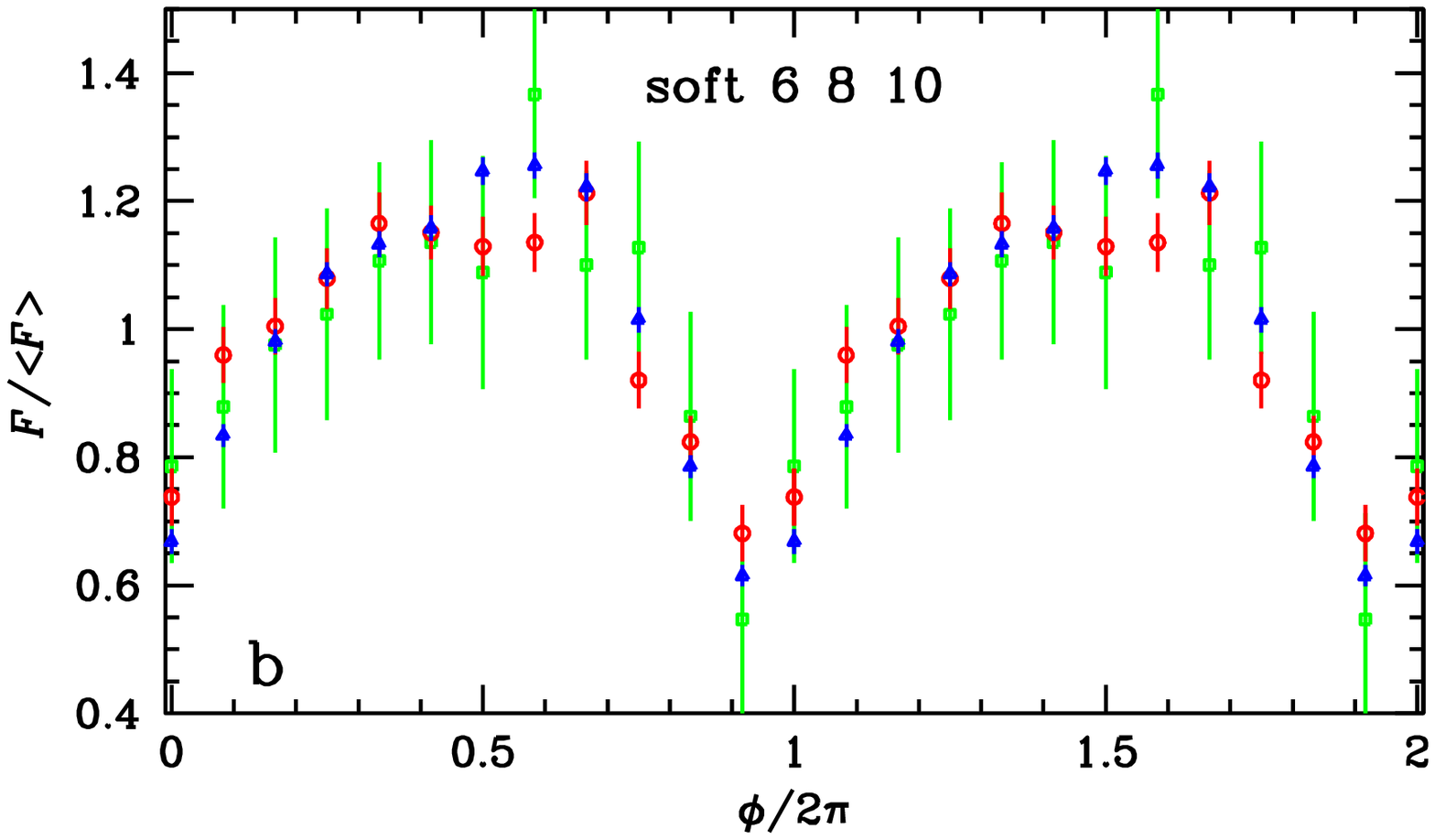}} 
\centerline{\includegraphics[width=\columnwidth]{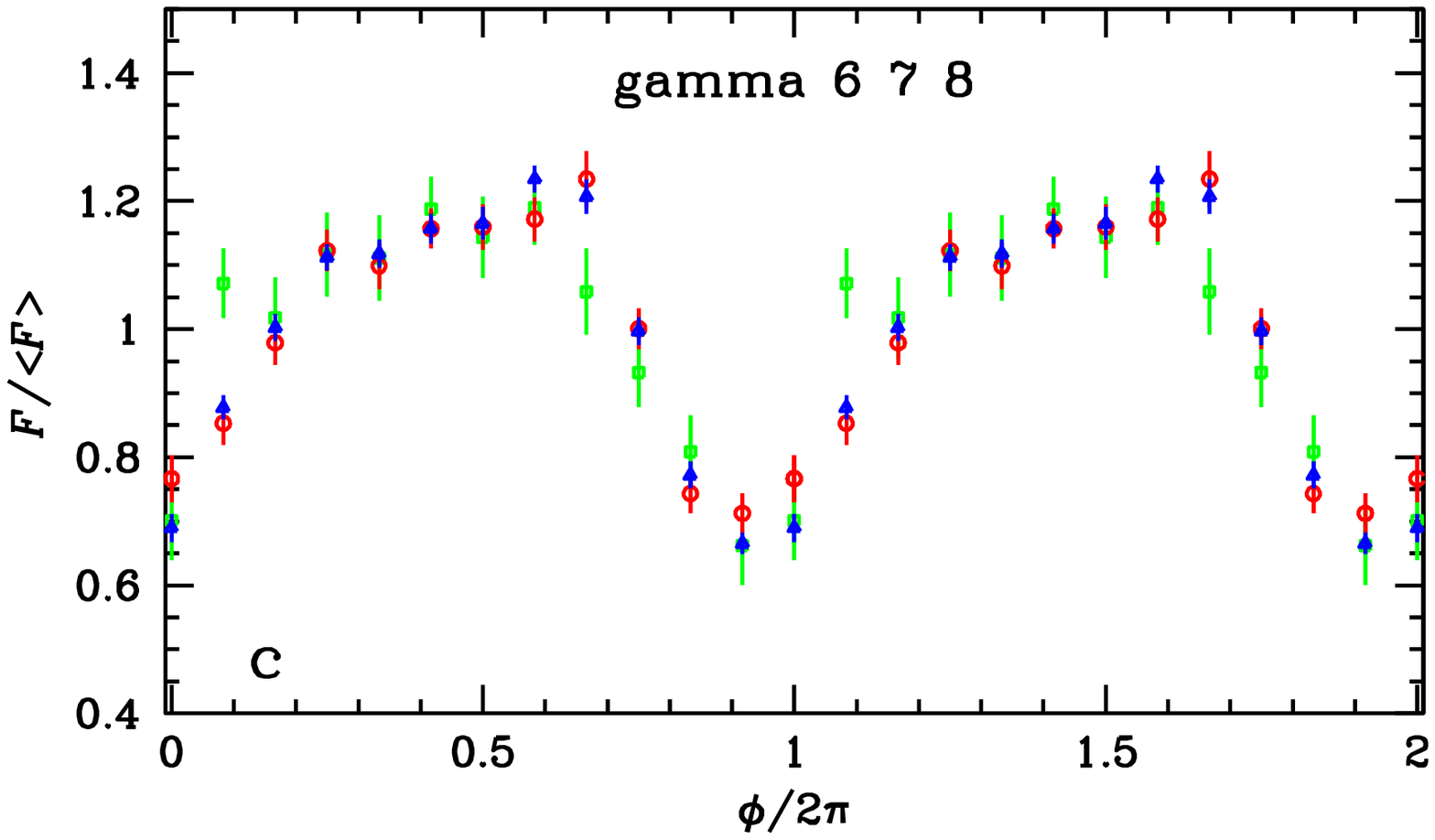}} 
\caption{Comparison of the folded light curves in different energy channels selected by spectral state. The blue, red and green light curves are for channels (a) 24--35 keV, 50--75 keV, 75--100 keV in the hard state, (b) 24--35 keV, 50--75 keV, 100--150 keV in the soft state, and (c) 24--35 keV, 35--50 keV, 50--75 keV during periods of high-energy \g-ray emission, respectively.
} \label{states}
\end{figure}

Fig.\ \ref{hard_soft}(a--b) compares the folded light curves for the count rates of the ASM and BAT instruments for the hard, intermediate and soft states, using the criteria described in Section \ref{data}, see also Fig.\ \ref{correlation}. We see that in the ASM range, the soft state modulation is significantly stronger than in the hard state, with the fractional modulation depth of $D\simeq 0.64$, 0.57, respectively. A similar result for the ASM data (but using different state criteria) is shown in fig.\ 6 of \citet{h08}. In the BAT range, the difference between the states is smaller, $D\simeq 0.49$ and 0.53 in the hard and soft state, respectively. The difference between the states may be due to the soft state taking place during periods of a higher wind mass loss rate. For the ASM and BAT, the ratio of the fitted values of $\tau_0$ of our wind model between the soft and hard states is $\simeq 1.20$ and $\simeq 1.08$, respectively. The latter value may be close to the ratio of the mass loss rates (as the dominant opacity in the BAT range is likely to be Compton), whereas the former is likely to be also affected by changing the bound-free opacity due to the change of the ionizing continuum.

We then compare the orbital modulation in different energy channels of the ASM and BAT, for all of the available data. We number the channels consecutively with the increasing energy, from 1 to 11. Fig.\ \ref{comparison} shows some examples of comparison of the modulation profiles in different energy channels, and its panels are marked with the channel numbers being compared. We see that although the modulation depth generally decreases with the increasing energy, the strongest modulation occurs for the 3--5 keV range, as shown in Fig.\ \ref{comparison}(a). We have checked that the same behaviour takes place within the hard and soft states. This effect appears to be due to an additional, less or differently modulated, spectral component appearing in soft X-rays, which is probably re-emission in soft X-ray lines of the absorbed continuum by the stellar wind, see SZ08. The general decrease appears then to be due to the bound-free absorption cross section decreasing (except for ionization edges) and the Compton scattering cross section decreasing as well with the increasing energy. The profiles also become more symmetric with the increasing energy, indicating that the electron density is more symmetric (with respect to the plane perpendicular to the binary plane along the line joining the stars) than the ionization structure. All the profiles have the minima within $\phi/2\upi\simeq 0.9$--1, with no indication of a different modulation pattern up to 150 keV. 

Fig.\ \ref{depth} shows the fractional modulation depth, $D$ of equation (\ref{eq:depth}), as a function of the photon energy, estimated using fits with the model of equations (\ref{tau_sc_v}), (\ref{flux}) with a free $\beta$ and of the 3-harmonic model. Since the obtained statistical errors are very small, a measure of the uncertainty is provided by the differences between the values for the two models. We show $D$ only up to 75 keV, as at higher energies strong noise prevents trustworthy estimates. We see that indeed the orbital modulation is the strongest in the 3--5 keV range, and decreases towards both lower and higher energies. Fig.\ \ref{depth} also shows $D$ from the ISGRI data (Section \ref{integral} below), which is in a good agreement with the BAT results. 

The dashed curve in Fig.\ \ref{depth} shows the dependence predicted using equation (\ref{eq:depth}) for Compton scattering alone, using the Klein-Nishina cross section and $\Delta \tau_{\rm T}=0.75$, for which the model dependence crosses the observed one around 20 keV. We see that the decline of $D$ above 5 keV cannot be accounted for by the decline of the Klein-Nishina cross section with energy, and photoionization absorption has to contribute at $\la 20$ keV (in agreement with, e.g., SZ08).

Figs.\ \ref{states}(a--b) show the modulation-profile dependence on energy during the hard and soft state, respectively. We see that up to $\simeq 100$--150 keV, at which energies the statistics become poor, the profiles have the minima around 0.9--1 phase, the same as at low energies and as for the entire data. An analogous figure for the intermediate state looks similar and it is not shown.

\begin{figure}
\centerline{\includegraphics[width=\columnwidth]{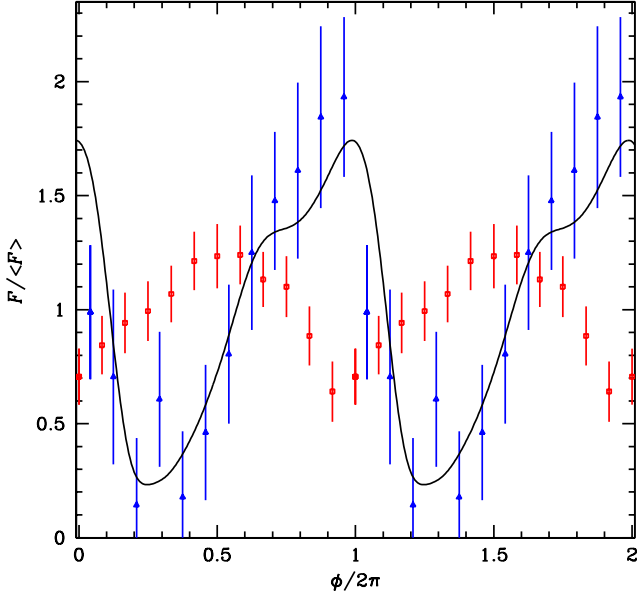}} 
\caption{Comparison of the folded light curves in the hard X-rays (red symbols) in the soft state in the 60--150 keV range with those at $\ge 0.1$ GeV (from FLC09, blue symbols). The constant background level (FLC09) has been subtracted from the LAT fluxes. The black curve shows the 100 keV modulation predicted (Z12) using the anisotropic Compton model fitted by \citet{dch10} to the GeV data. All the profiles are normalized by the respective average values.
} \label{fermi_hardX}
\end{figure}

Fig.\ \ref{states}(c) shows the modulation for the data simultaneous with the observations of high-energy \g-ray emission (see Section \ref{data}). We see the same modulation pattern as for other data, and, in particular, neither a change of the positions of the maximum and minimum nor a flattening of the profile with respect to the profiles at other states up to at least 75 keV. At higher energies, we also see no indication of the presence of another modulation pattern, but the data become noisy. 

The blue symbols in Fig.\ \ref{fermi_hardX} show the orbital modulation observed at $E\ge 0.1$ GeV by the \fermi\/ LAT (FLC09). The black curve shows the modulation of the jet emission predicted at 100 keV using the anisotropic Compton model \citep{dch10}, from fig.\ 6 in Z12. The 100 keV model is relatively similar to that at 0.2 GeV, though somewhat shifted to higher phases (Z12). These jet modulation profiles are shifted in phase with respect to those observed in X-rays, and have the maximum around the superior conjunction. We compare then the predicted modulation of the jet spectral component with that observed at 60--150 keV, red symbols. This profile has been obtained by averaging the BAT 100--150 keV profile with the \integral\/ one for 60--100 keV (see Section \ref{integral} below), both in the soft state. Indeed, we see it is similar to the modulation profiles observed in other states and very different from either the jet contribution predicted at 100 keV or observed at $>0.1$ GeV. 

We note that \citet{dch10} have determined the location of the \g-ray emitting region at a distance $\sim 2 a$ along the jet originating at the compact object. Then, X-ray emission from the jet would be much less absorbed by the wind than that originating at the compact object. Furthermore, if the jet X-ray emission were stronger in the soft state then the modulation amplitude would then decrease. Instead, we see in Fig.\ \ref{states} that the modulation depth either in the soft state or for the data simultaneous with the observed \g-ray emission is basically the same as in the hard state, during which there is no detectable jet contribution. We note that the modulation pattern of the jet 100 keV emission can be flatter than that shown in Fig.\ \ref{fermi_hardX} due to the electron cooling slower than at $>0.1$ GeV, causing the 100-keV emission region along the jet to be more extended than that in high-energy \g-rays. Still, a major jet contribution would flatten the observed total modulation profiles, which is not seen. Thus, based on Figs.\ \ref{states}--\ref{fermi_hardX} (also Section \ref{integral} below), we see no indication of a contribution from the low-energy tail of the \g-ray emission. That emission has to have a low-energy cutoff above the hard X-ray range. Indeed, the jet models with the minimum Lorentz factors of the accelerated electrons of 1300--1500, which are shown by the blue curves in figs.\ 5(a--b) in Z12, do satisfy this constraint and contribute very little at 100 keV. On the other hand, their models with the minimum Lorentz factors of 300 and 700 predict too much flux at 100 keV and appear to be ruled out.

\section{Orbital modulation in \textit{INTEGRAL} data}
\label{integral}

\begin{table}
\centering \caption{The log of the \integral\/ ISGRI observations. The total exposure is about $4.5\times 10^6$ s. The first row includes all the observations before the start of the BAT monitoring. The low 14--50 keV state corresponds to the soft state as defined in Section \ref{monitoring}. 
}
\begin{tabular}{c  c  c c}
\hline
Start [MJD] & End [MJD] & Exposure [s] & 14--50 keV state \\
\hline
52629.632 & 53353.532 & 2030100 & mixed \\
53365.239 & 53745.081 &   32748 & high \\
53750.259 & 53881.889 &  152052 & low \\
53899.027 & 53912.449 &    4349 & high \\
53912.449 & 54059.913 &  402936 & low \\
54241.896 & 54241.986 &    6586 & low \\
54438.871 & 54444.899 &  126269 & high \\
54445.077 & 54453.108 &  168612 & low \\
54574.516 & 54609.314 &  222279 & low \\
54771.169 & 54794.013 &  341155 & low \\
54824.233 & 54973.407 &  277455 & high \\
55179.050 & 55186.242 &   69941 & high \\
55341.028 & 55341.047 &    1796 & low \\
55500.860 & 55542.037 &  651859 & high \\
\hline
\end{tabular}
\label{isgri_log}
\end{table}

\begin{figure}
\centerline{\includegraphics[width=\columnwidth]{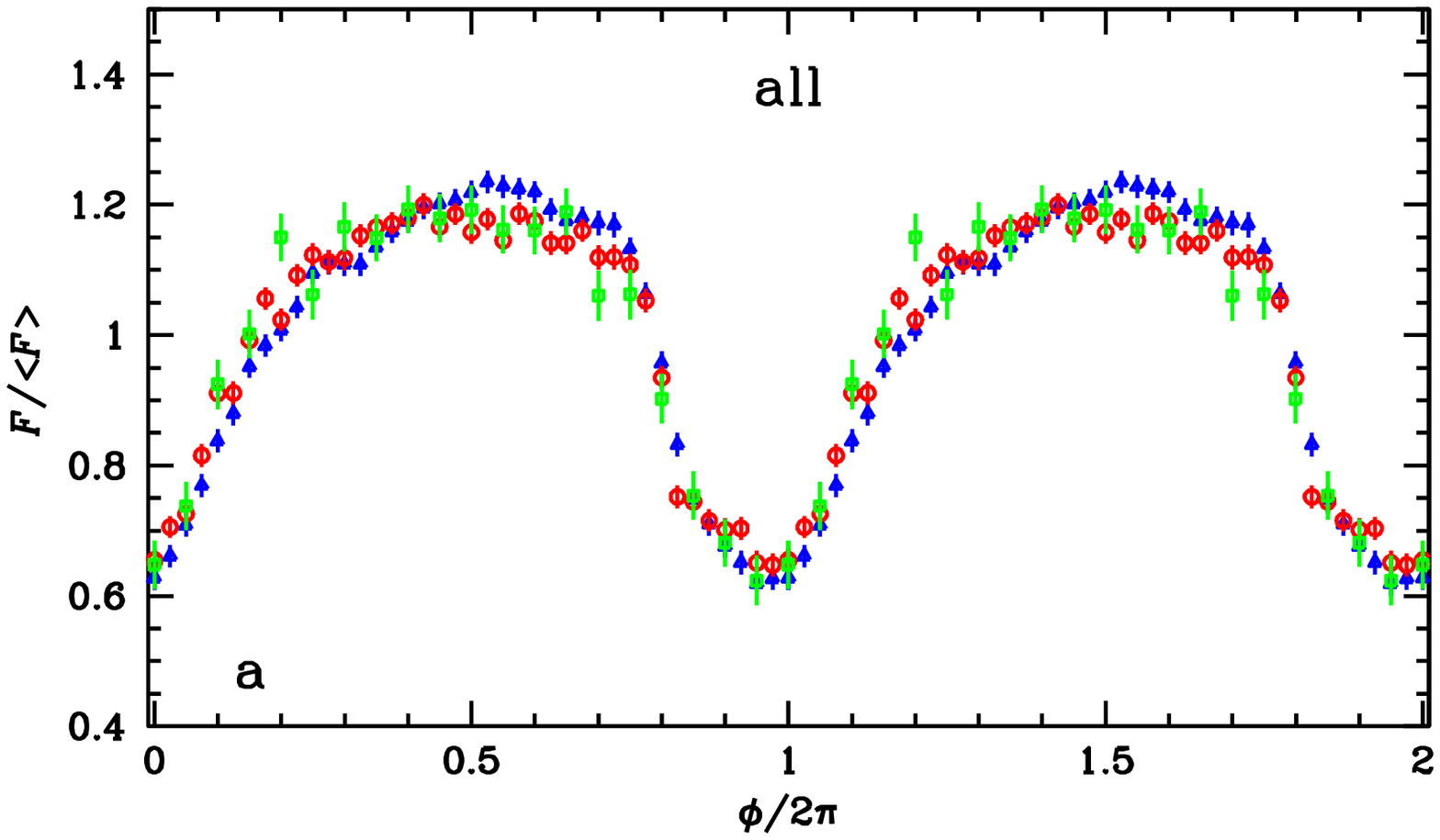}} 
\centerline{\includegraphics[width=\columnwidth]{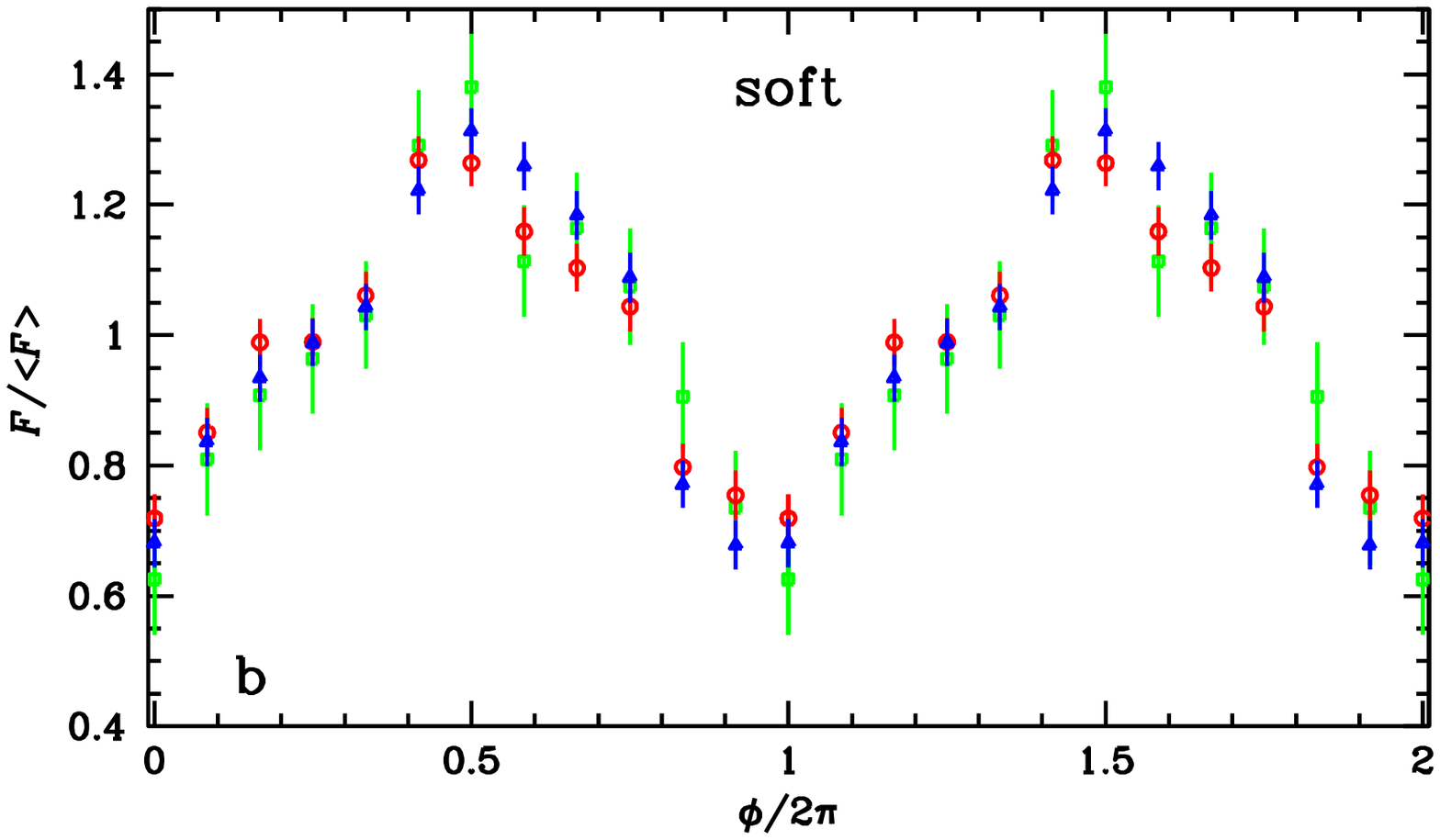}} 
\caption{Folded light curves for \integral/ISGRI observations, taking into account (a) all observations and (b) those corresponding to the soft state. The blue, red and green light curves are for energy ranges of 20--40 keV, 40--60 keV, 60--100 keV, respectively. 
} \label{int_profiles}
\end{figure}

We have studied \integral\/ data from the ISGRI detector (a part of the IBIS
telescope, \citealt{ubertini03}). We have used all the data public as of 2012 May 1, available at the \integral\/ Science Data Centre (ISDC). We select only data with the off-axis angle $< 5\degr$, which corresponds to the fully coded part of the detector (the inner $10\degr \times 10\degr$ of the field of view). These data span MJD 52629--55542. \integral\/ observations consist of pointings with a typical length of 2--3 ks. Our data set comprises of 1824 such
pointings, for a total exposure of $\simeq 4.5 \times 10^6$ s. We consider only the 20--100 keV energy range since the data above 100 keV have relatively poor statistics. The data have been reduced using the Offline Scientific Analysis v.\ 9.0 package provided by the ISDC \citep{courvoisier03}, with the pipeline parameters set to the default values. The created ISGRI light curves have a resolution of 100 s. Table \ref{isgri_log} gives the log of the observations.

We consider first all the observations, and then only those corresponding to the soft-state intervals with the low BAT flux, see Section \ref{data}. We perform the analysis using the standard {\tt ftools} software, in which the folded light curves are unweighted averages of the fluxes within individual bins. Our results are presented in Fig.\ \ref{int_profiles}. We see no indication of a change of the profile shape towards high energies in either of the two cases considered, confirming our results in Section \ref{lc} above. We have also calculated the modulation profiles for only the data simultaneous with \g-ray detections. We have obtained results entirely consistent with those for the soft state, but with somewhat worse statistics.

In Fig.\ \ref{asm_bat_integral}(a), we compare the 20--40 keV modulation profile for all the ISGRI data with the corresponding BAT profile. We see a very good agreement between them, which is consistent with the mostly overlapping energy coverage of the analyzed data. A 20--40 keV modulation profile for an early observation of Cyg X-3 by ISGRI is presented in \citet{v03}, whose results are similar to ours but with substantially higher statistical fluctuations. That paper also presents the modulation profile from the monitoring by the \gro/BATSE during 1991--2000, which is also of almost the same shape as those of our ISGRI and BAT profiles. 

\section{\textit{RXTE}\/ PCA/HEXTE phase-resolved spectra}
\label{xte}

Here, we study the \xte\/ Proportional Counter Array (PCA) and High Energy X-ray Transient Experiment (HEXTE) data sets used by SZM08, who obtained average X-ray spectra in five spectral states. The log of the observations is given in \citet{h09}. We split each average data set into two subsets corresponding to the superior and inferior conjunctions, which is done by dividing the orbital phases into two parts, $\phi/2\upi=0.28$--0.78 and 0.78--0.28. These boundaries have been determined using the orbital template of \citet{vb89} with the condition that the orbital period is divided into two parts of equal length, with the template flux greater or lower than that corresponding to the dividing phases.

We have performed standard \xte\/ PCA and HEXTE data reduction using {\tt ftools}. The resulting exposures for the two subsets of the data of SZM08 are given in Table \ref{xte_log}. We see that whereas the states 1--3 have exposures $\ga 10^4$ s for each subset for either PCA or HEXTE, the states 4--5 have much worse coverage, especially for the PCA, with the exposures as short as $\simeq 1000$ s and $\simeq 3000$ s. Therefore, we present only the spectral shapes for the states 4--5, whereas we study in detail the differences between the subsets around the superior and inferior conjunctions for the states 1--3. 

\begin{table}
\centering \caption{The log of the \xte\/ observations. The exposures are for the sums of all the used observations within each phase interval within a state.
}
\begin{tabular}{c c  c  c c}
\hline
Spectral state &  \multicolumn{2}{c}{Inferior Conjunction} & \multicolumn{2}{c}{Superior Conjunction}\\
&\multicolumn{2}{c}{Exposure [s]} &\multicolumn{2}{c}{Exposure [s]}\\
 & PCA & HEXTE & PCA & HEXTE \\
 \hline
1 & 9120 & 9120 & 23408 & 19549 \\
2 & 24048 & 14611 & 42976 & 25952 \\
3 & 20912 & 23049 & 23008 & 9337 \\
4 & 3744 & 4645 & 2960 & 3805 \\
5 & 1120 & 5465 & 12688 & 11672 \\
 \hline
\end{tabular}
\label{xte_log}
\end{table}

In our approach, we study the difference between the spectra around the inferior and superior conjunctions assuming it is due to an additional optical depth, with scattering and absorption characterized by a constant ionization coefficient, see equation (B3) of \citet{zdz12}. We first fit the spectra around the inferior conjunction. Finding a physical model of unabsorbed spectra of Cyg X-3 has proven, in this and previous studies, to be difficult. For example, \citet{h08,h09} found that fitting data at $\ga 3$ keV do not allow an unambiguous determination of the intrinsic spectra, with the range of possible models strongly varying at low energies. On the other hand, Cyg X-3 data from \sax\/ extend to $<1$ keV, and are better suited for such a determination. Those data have been studied by SZ08, who used a relatively complex model of wind absorption and emission, modifying spectra from hybrid plasma and Compton reflection \citep{pc98,c99,gierlinski99}.

\begin{figure}
\centerline{\includegraphics[width=6.1cm]{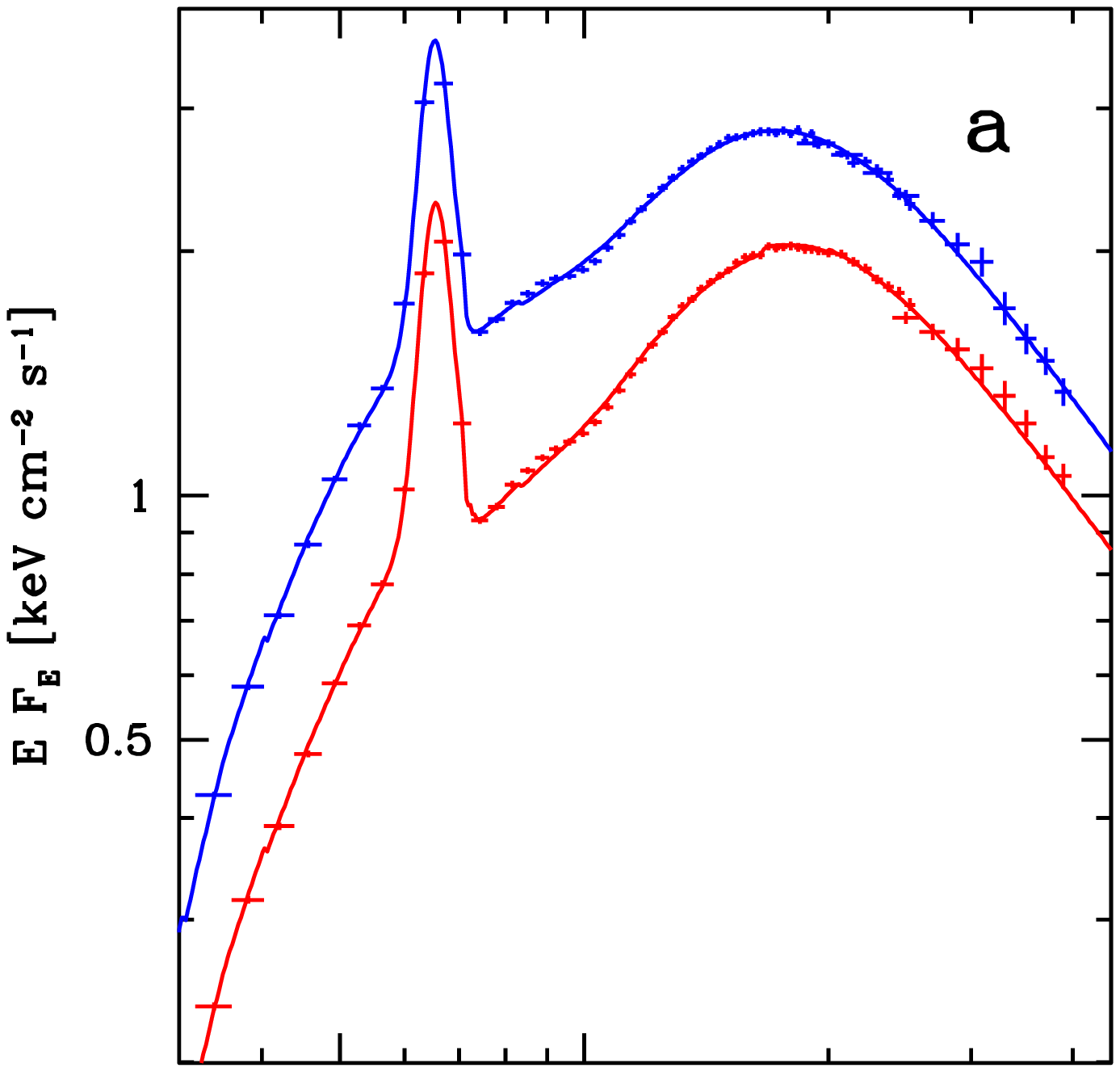}} 
\centerline{\includegraphics[width=6.1cm]{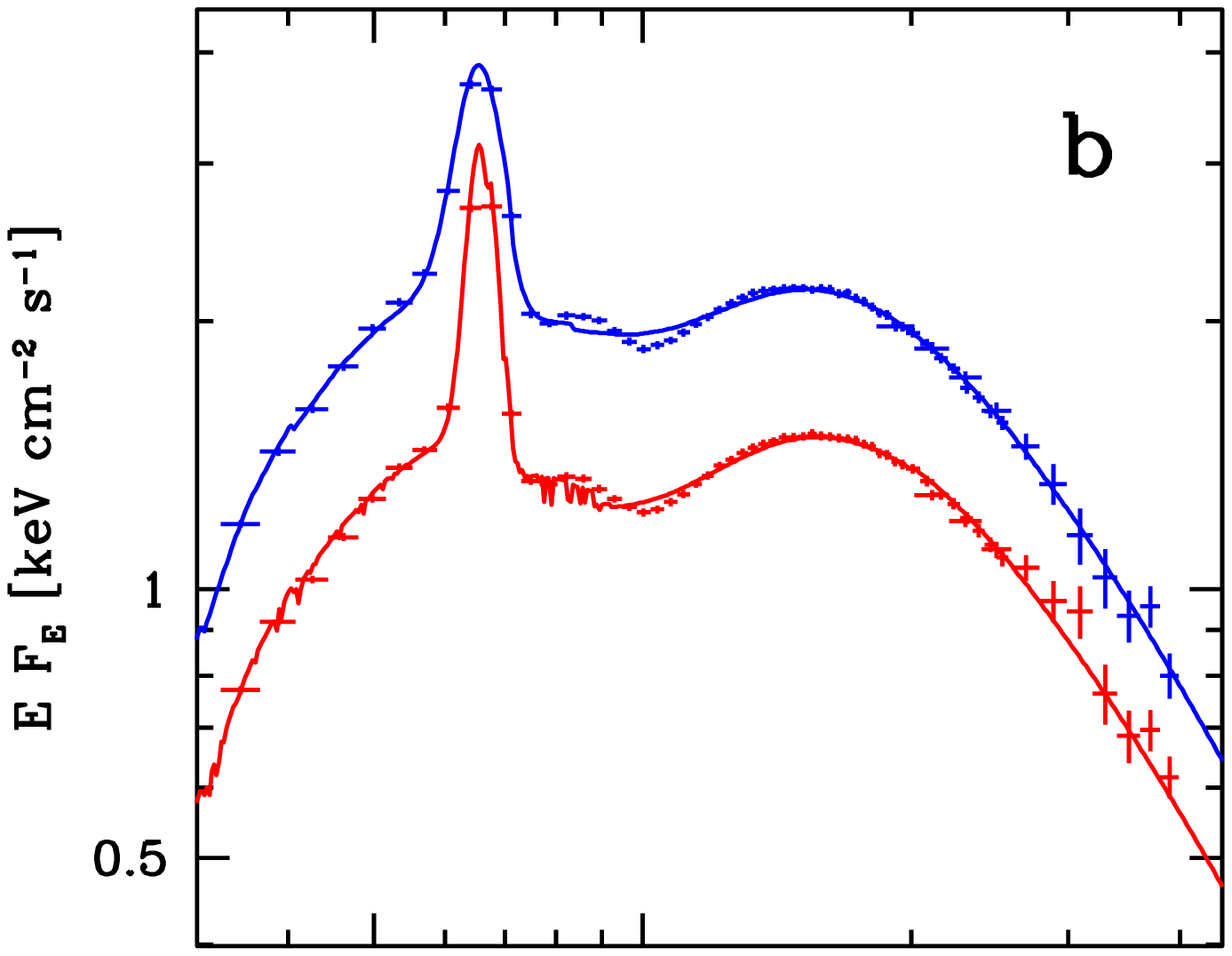}} 
\centerline{\includegraphics[width=6.1cm]{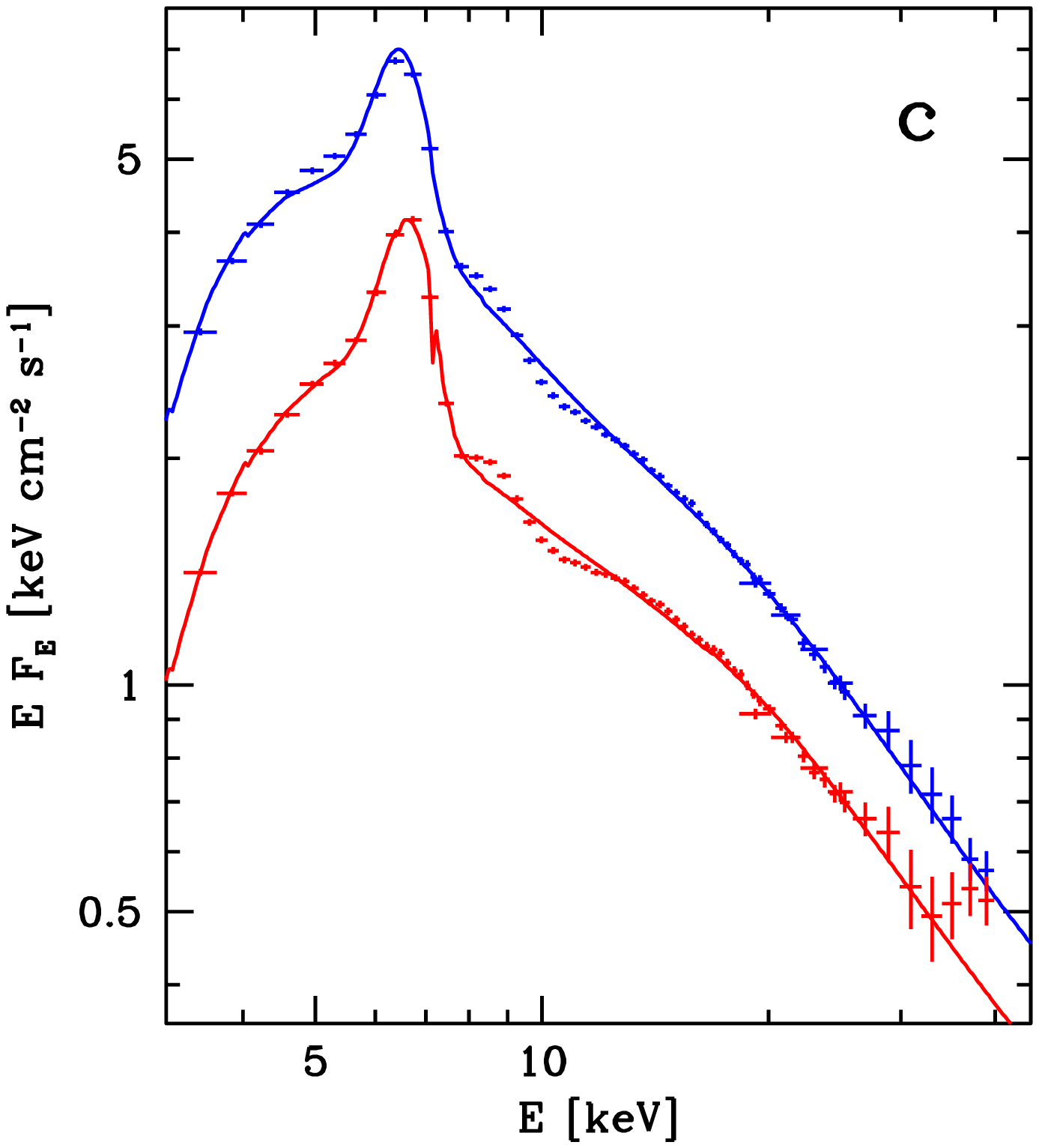}} 
\caption{The panels (a--c) show the spectra for the states 1--3, respectively, in the classification of SZM08. The upper (blue) and lower (red) spectra correspond to the phase ranges around the inferior and superior conjunction, respectively. The crosses show the unfolded PCA+HEXTE spectra. The inferior spectra are modelled by emission of hybrid plasma and reflection/Fe K line absorbed by a dual absorber. The models of the superior spectra are the best-fit models of the inferior spectra absorbed/scattered in an additional medium, corresponding to the difference in the average column density between the two phase ranges.
} \label{f:spectra}
\end{figure}

\begin{figure}
\centerline{\includegraphics[width=5.3cm]{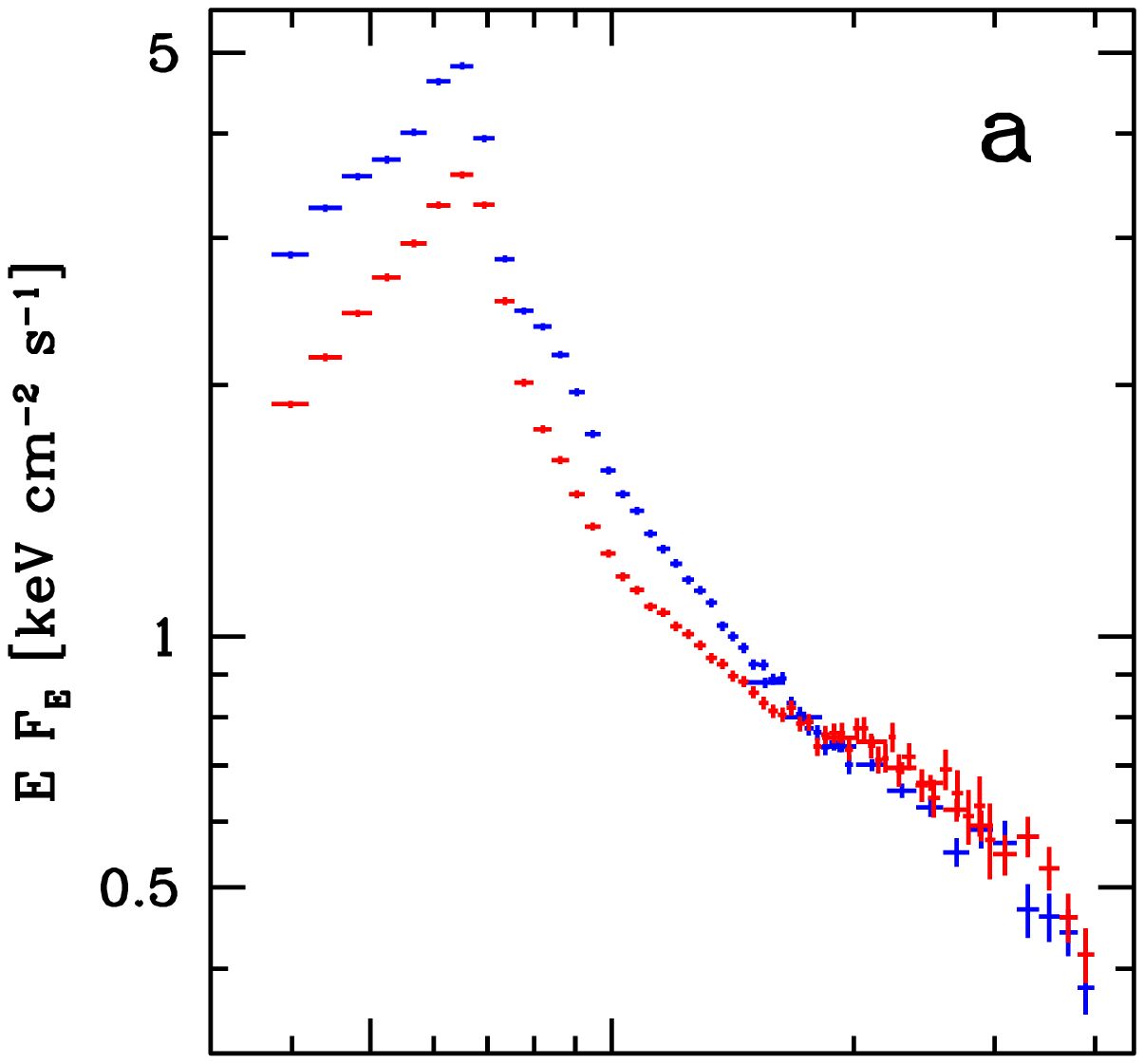}} 
\centerline{\includegraphics[width=5.3cm]{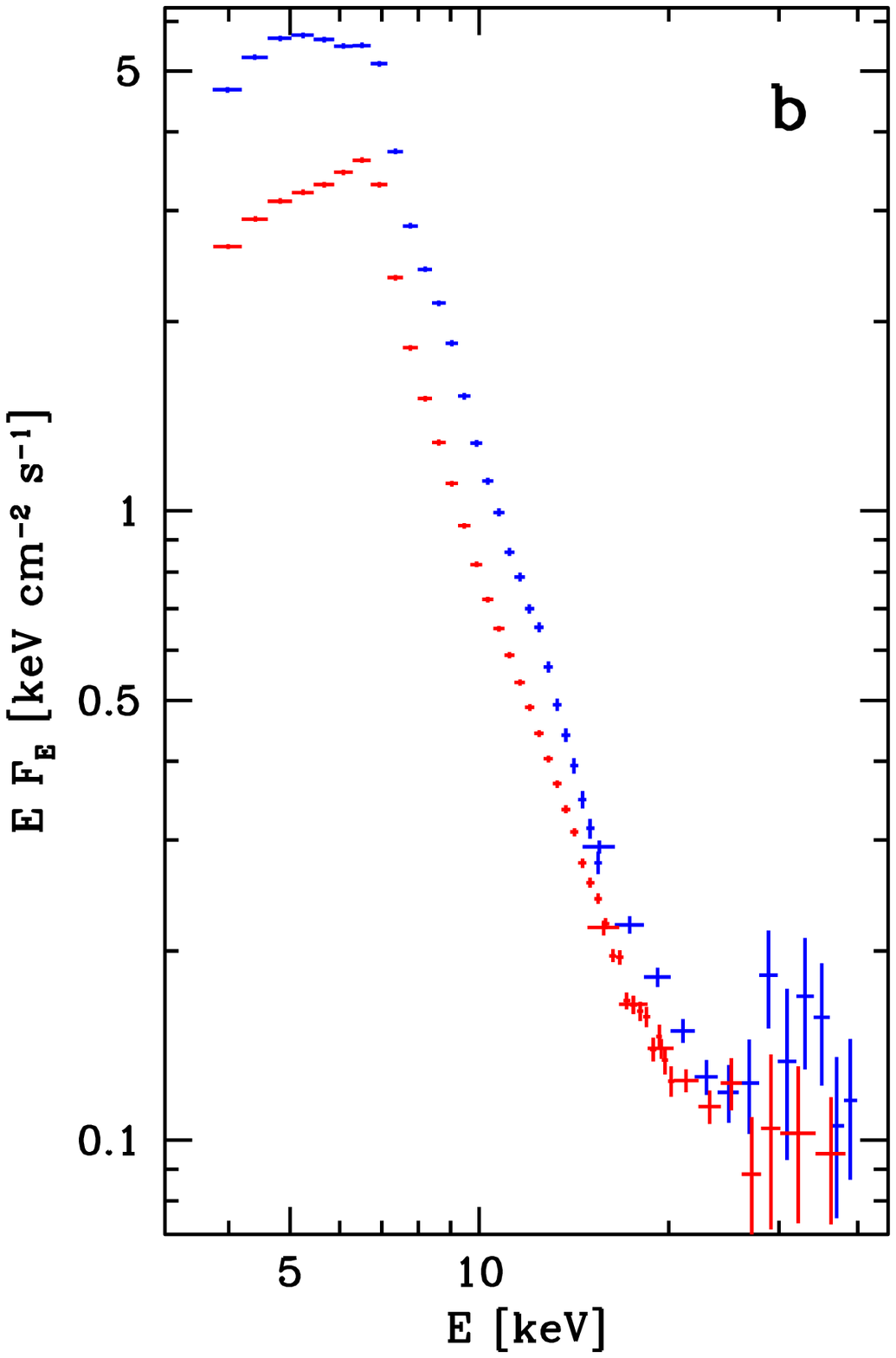}} 
\caption{The spectra of the states 4-5, shown in the panels (a--b), respectively. The upper (blue) and lower (red) spectra correspond to the phases around the inferior conjunction and superior conjunction, respectively. The crosses represent the unfolded PCA+HEXTE spectra. The models (not shown) are given by emission of hybrid plasma and reflection/Fe K line absorbed by a dual absorber. 
} \label{f:45}
\end{figure}

Here, we fit the data from PCA and HEXTE, which are for $E\ga 3$ keV only, and do not allow unambiguous determination of the low-energy intrinsic spectra. However, our objective is to study physical mechanisms of the orbital modulation, and determining the actual intrinsic spectral distribution is beyond the scope of this work. We thus fit the data with a model that has been shown to provide relatively good fits to a large set of data from \xte\/ and \integral\/ (\citealt{v03,h08,h09}, SZM08, \citealt{corbel12}), but which still should be considered as phenomenological. In particular, any inferences about the unabsorbed spectra based on that model are very uncertain. The model is based on emission of hybrid plasma and reflection (including the Fe K line), similar to the model used by SZ08. However, the complex absorption by the wind is treated only phenomenologically, as a product of full and partial neutral absorbers. The Fe abundance of the absorber and reflector is assumed to be free. However, given the neglect of photoionization and the complex spatial structure of the wind, the fitted value should also be considered as phenomenological only. In spectral fitting, we allow a free normalization of each HEXTE spectrum with respect to the corresponding PCA one, and we add a 1 per cent systematic error to each of the count spectra. The PCA and HEXTE data are fitted in the energy ranges of about 3--25 keV and 20--40 keV, respectively (as we have found the statistical quality of the HEXTE spectra above 40 keV is relatively poor). The resulting fits to the spectra in the states 1--3 are statistically satisfactory, $\chi^2/\nu\la 1$. The fitted parameters are consistent with those presented in \citet{v03} and \citet{h08,h09}.

Then, we take the best-fit model to an inferior-conjunction spectrum, and fit the corresponding superior-conjunction one with the inferior-conjunction spectral model but attenuated by absorption and scattering in an additional medium due to lines of sight through the wind around the superior conjunction being longer than those around the inferior conjunction. We have found that the fractional modulation depth averaged over our phase intervals is $D\simeq 0.4$ but it strongly decreases at $E\ga 15$ keV. We have first tested whether this behaviour can be due to the Klein-Nishina decline of the Compton cross section. However, that decline is much too slow to account for the observed energy-dependent modulation depth, which confirms our results in Section \ref{lc} above. Clearly, photoionization, known to be present in Cyg X-3 (e.g., SZ08), also contributes. Its cross section generally declines fast with increasing energy (apart from ionization edges). Accordingly, we have found that good fits to the data can be obtained including absorption by an ionized medium. Specifically, our model for the additional superior-conjunction attenuation consists of partial covering by a partially ionized medium and full covering by a fully ionized medium. Such a model may approximate shadowing of the stellar wind by the X-ray source, and it has been invoked to explain the IR modulation and spectral lines \citep{v93}. The attenuation is then by the factor,
\begin{equation}
A(E)=\exp(-N_{\rm full}\sigma_{\rm C})\left\{1-f_{\rm cov}+f_{\rm cov}\exp\left[-N_{\rm part}\left(\sigma_{\rm C}+\sigma_{\rm ion}\right)\right]\right\},
\label{eq:abs}
\end{equation}
where $N_{\rm full}$ and $N_{\rm part}$ are the electron column densities of the fully and partially covering medium, and $f_{\rm cov}$ is the covering fraction. We use the ionized absorber of \citet{reeves08}, based on {\tt xstar} \citep{bk01}. It is characterized by the column density and ionization parameter, $\xi\equiv L_{\rm X}/(n r^2)$, where $n$ is the medium density and $r$ is the distance from the ionizing source to the medium; see \citet{reeves08} for further details. Our model is clearly very simplified. We have neglected spatial variations of the ionization coefficient. Also, we have assumed that scattered photons are entirely removed from the line of sight, whereas some photons are also scattered into the line of sight from other directions. The fit results are summarized in Table \ref{t:abs}, and the unfolded spectra and the models are shown in Fig.\ \ref{f:spectra}.

\begin{table}
\centering \caption{The fitted parameters of the additional absorbing medium related to the superior conjunction for the states 1--3 (which have the exposures sufficient for averaging). See Section \ref{xte} for details.
}
\begin{tabular}{c c  c  c c c}
\hline
State & $N_{\rm full}$ & $N_{\rm part}$ & $f_{\rm cov}$ & $\log_{10}\xi$ & $\chi^2\!/\nu$\\
& $10^{23}\,{\rm cm}^2$ & $10^{23}\,{\rm cm}^2$ &  & erg cm s$^{-1}$ \\
 \hline
1 & $2.0\pm 0.2$ & $6.8\pm 0.5$ & $0.36\pm 0.01$ & $1.6\pm 0.4$ & 40/53\\
2 & $2.1\pm 1.4$ & $10.8\pm 8.7$ & $0.31\pm 0.08$ & $3.3\pm 0.4$ & 30/55\\
3 & $3.5\pm 0.3$ & $3.0\pm 0.6$ & $0.44\pm 0.07$ & $1.8\pm 0.4$ & 55/56\\
 \hline
\end{tabular}
\label{t:abs}
\end{table}

We see that the total electron column density is $N\sim 10^{24}$ cm$^2$, which corresponds to the difference in the average Thomson optical depth between the two phase intervals of $\Delta \tau_{\rm T}\sim 1$. This is in agreement with $\Delta\tau\sim 1$ found from fitting the orbital modulation profiles (Section \ref{lc}). The fitted values of $\xi\sim 10^{2-3}$ erg cm s$^{-1}$ correspond to moderate ionization. We can notice that the models do not fit the smeared edges clearly seen around $\sim 10$ keV, which are apparently due to absorption by H-like Fe in a very strongly (but not fully) ionized phase of the wind, which we do not model here. Using the definition of $\xi$, the characteristic distance of the ionized medium is $r\sim L_{\rm X}/(N \xi)$, which, for the fitted values of $\xi$ and $N$ is $r \sim 10^{11-12}$ cm. This is of the order of the orbital separation, $a$, as expected. Thus, the fitted model is consistent with bound-free absorption taking place in the stellar wind. We note that the fitted parameters do not form monotonic sequences with the increasing state number. This appears to be due to the limited observation exposures. Then, the two sub-states of a given state are not true averages, but are affected by variability taking place during the used observations. 

In Fig.\ \ref{f:45}, we show the inferior and superior spectra for the states 4--5. We can see that due to the short exposures, the intrinsic variability is apparently comparable to the orbital modulation, especially for the spectra of the state 4, which have the superior-phase fluxes at $\ga 20$ keV higher than those of the inferior phases, contrary to the expectations for long-term averages.

\section{Conclusions}
\label{conclusions}

We have studied orbital modulation of X-rays from Cyg X-3. Using the data from the detectors ASM, PCA and HEXTE of \xte, BAT of \swift, and ISGRI of \integral, we have found that the modulation depth at $E\ga 5$ keV decreases with the photon energy. The decrease is too fast to be accounted for by the Klein-Nishina decrease of the Compton cross section with energy. As we confirm by fitting the PCA/HEXTE data, the energy dependence requires the presence of a moderately ionized absorber, with the ionization coefficient of $L_{\rm X}/(n r^2)\sim 10^{2-3}$ erg cm s$^{-1}$. We have also found a decrease of the depth of the modulation at $E\la 3$ keV, which is probably due to the re-emission of the absorbed continuum in soft X-ray lines by the wind.

Given that we use the data accumulated over long time, we determine the energy-dependent folded and averaged light curves with high accuracy. The modulation profiles are not consistent with a spherically symmetric wind. The measured detailed shapes of the orbital modulation and their photon-energy dependence can serve for determination of the wind structure in future work. Also, the energy dependence of the modulation depth at low energies can be further studied using spectra from detectors sensitive at soft X-rays.

Assuming that the modulation is from wind absorption, the phase of the flux minimum corresponds to the maximum optical depth through the wind, which is then close to the superior conjunction. On the other hand, high-energy \g-rays detected from Cyg X-3 by \fermi\/ are found to have an orbital modulation shifted by about a half of the period with respect to that of the X-rays (FLC09). This pattern has been interpreted as due to anisotropy of scattering of stellar photons by relativistic electrons in the jet by \citet{dch10}. As calculated by Z12, the jet emission at hard X-rays has the modulation pattern almost the same as that in the GeV range. Thus, the jet X-ray contribution can potentially be measured by determining the orbital-modulation profiles at hard X-rays. 

We have performed such study, and have found no presence of an X-ray modulation pattern different from the standard one, i.e., with the minimum around the superior conjunction, up to $\sim$100 keV. This implies that an X-ray jet contribution is weak up to at least that energy and the jet component has to have a low-energy cutoff. This is consistent with the X-ray spectral measurements taken during periods of \g-ray emission, which do not show a hint of a jet component up to comparable energies (Z12; \citealt{corbel12}). This constraint is satisfied by two among the models presented in Z12, those with the minimum Lorentz factors of the accelerated electrons of 1300 and 1500.

\section*{ACKNOWLEDGMENTS}

This research has been supported in part by the Polish NCN grants N N203 581240, N N203 404939, 362/1/N-INTEGRAL/2008/09/0. We thank P. Lubi{\'n}ski for help with the analysis of the BAT data, S. Kitamoto for providing us with his updated ephemeris, and G. Dubus, the referee, for valuable suggestions and comments. We acknowledge the use of quick-look results provided by the \xte\/ ASM team, and of data obtained through the HEASARC online service provided by NASA/GSFC.

\label{lastpage}
\end{document}